\documentclass{iopjournal}
\usepackage{graphicx}
\usepackage{dcolumn}
\usepackage{bm}

\usepackage[utf8]{inputenc}
\usepackage[T1]{fontenc}
\usepackage{mathptmx}
\usepackage{amsmath}
\usepackage{etoolbox}
\usepackage{multirow}
\usepackage{booktabs}
\usepackage{xcolor}
\usepackage[percent]{overpic}
\usepackage{ulem}

\makeatletter
\def\@email#1#2{%
 \endgroup
 \patchcmd{\titleblock@produce}
  {\frontmatter@RRAPformat}
  {\frontmatter@RRAPformat{\produce@RRAP{*#1\href{mailto:#2}{#2}}}\frontmatter@RRAPformat}
  {}{}
}%
\makeatother

\begin{document}

\articletype{Paper} 

\title{Gyrokinetic global simulation of Alfv\'enic ion temperature gradient mode in reversed magnetic shear}

\author{Gengxian Li$^{1,*}$, Zhixin Lu$^{1,*}$, Philipp Lauber$^1$, Matthias Hoelzl$^1$, Guo Meng$^1$ and Yong Xiao$^2$}

\affil{$^1$Max Planck Institute for Plasma Physics, Boltzmannstr. 2, Garching, 85748, Germany}

\affil{$^2$Institute for Fusion Theory and Simulation, Zhejiang University, Hangzhou, China}

\affil{$^*$Author to whom any correspondence should be addressed.}

\email{Gengxian.Li@ipp.mpg.de}
\email{Zhixin.Lu@ipp.mpg.de}


\begin{abstract}
In this work, a systematic study of electromagnetic instabilities driven by the temperature gradient in magnetically confined fusion plasmas with reversed magnetic shear is conducted using gyrokinetic particle-in-cell simulations. An electromagnetic instability arising in the low-$\beta$ regime is investigated, where $\beta=8\pi nT/B^2$ denotes the ratio of plasma pressure to magnetic pressure. Within a reversed shear safety factor ($q$) profile, when a mode rational surface coincides with the position of zero shear, an instability dominated by only one poloidal harmonic emerges, rather than the conventional ion-temperature-gradient (ITG) mode. Simulation results demonstrate that the instability exhibits pronounced electromagnetic polarization even in the low-$\beta$ regime, with a real frequency significantly higher than that of ITG modes, and show that it is destabilized by the temperature gradient and not by the density gradient. This instability can be observed even for a monotonic $q$ profile with weak magnetic shear. Based on a systematic comparison with other typical electrostatic and electromagnetic instabilities, this instability is identified as a weak shear Alfv\'enic-ion-temperature-gradient (\textit{WSAITG}) mode, which may provide an explanation for the low-frequency Alfv\'en modes (LFAM) observed in experiments. Wave–particle resonance analysis in phase space reveals that, in contrast to the ITG mode, well-passing particles provide an additional resonant population that drives the WSAITG mode.
\end{abstract}

\section{\label{sec:introduction}Introduction}
Modern tokamak research aims to develop stable and high-performance operational regimes. One class of advanced scenarios is achieved by tailoring the bootstrap current fraction to modify the safety factor ($q$) profile in the plasma core region\cite{RJGoldston_1994}. This control can lead to reduced or negative magnetic shear in the core region. In reversed magnetic shear configurations, the confinement in the core region is significantly improved, which corresponds to the formation of the internal transport barrier (ITB) \cite{connor2004review, wolf2002internal, PhysRevLett.132.065106, mysq-8cv3}. Such operational regimes have been widely implemented in major tokamak experimental devices \cite{PhysRevLett.72.3662, Yu_2016, PhysRevLett.75.4417,PhysRevLett.78.2377,Gong_2024}, highlighting the importance of understanding the underlying instabilities near the transport barrier. Previous global gyrokinetic simulations have shown that, in this region with zero magnetic shear, a mode transition from ion temperature gradient (ITG) mode to the kinetic infernal mode (KIM) occurs as the plasma $\beta$, the ratio of thermal pressure to magnetic pressure, increases \cite{gxli_2025, 10.1063/5.0013349}. However, in reversed-shear configurations, the radial distribution of rational surfaces is highly sensitive to the minimum value of the $q$ profile \cite{10.1063/1.3243918}, $q_{\min}$. The mode transition under different characteristic rational surface distributions has received little attention so far.

In tokamak plasmas, instabilities located on the Alfv\'en continuous spectrum are difficult to excite due to continuum damping \cite{10.1063/1.2838239,RevModPhyschen}. Previous theoretical studies have suggested that the Alfv\'en continuous spectrum could be destabilized by a finite ion temperature gradient \cite{FulvioZonca_1996}. In this case, finite Larmor radius (FLR) and finite orbit width (FOW) effects lead to the discretization of the unstable continuum. As a result, a distinct instability Alfv\'enic-ion-temperature-gradient (AITG) mode can arise at plasma $\beta$ values well below the ideal MHD ballooning threshold \cite{FulvioZonca_1998}. In earlier studies, AITG mode and kinetic ballooning mode (KBM) are often referred to as the same type of instability. There have been relatively few experimental \cite{Chen_2016} and simulation \cite{GZhao2002pop} studies on the AITG mode. In the experimental report of AITG mode, the equilibrium is characterized by reversed magnetic shear \cite{Chen_2016}.  In such configurations, the conventional ballooning transformation is no longer applicable as the magnetic shear vanishes, making theoretical investigation more complicated. Besides, owing to the conditions required for its excitation, the AITG mode is generally believed to have a significant impact on both ion and electron transport, highlighting the importance of understanding the underlying physics and characteristics of the AITG mode. Therefore, we perform global gyrokinetic electromagnetic simulations using the TRIMEG-GKX code \cite{lu2019development,lu2026cpc} to investigate this instability.

In this work, we first identify different instabilities under two slightly different $q$ profiles and compare their mode structures as well as real frequencies. It is found that in the low to moderate $\beta$ regime, when the position of $q_{\min}$ coincides with the location of a mode rational surface, an instability dominated by one single poloidal harmonic ($m$) emerges, rather than the conventional ITG mode. This instability exhibits a dispersion relation similar to the ITG mode, but with a significantly higher real frequency. Polarization analysis indicates that this instability is electromagnetic. Meanwhile, it can be driven exclusively by the temperature gradient, in contrast to the KBM, which is driven by the pressure gradient. By additional analysis of the mode characteristics, this instability is identified as an AITG mode. It may also provide a plausible explanation for the low frequency Alfv\'en modes (LFAM) observed in experiments\cite{Heidbrink_2021,10.1063/5.0141186}. Its excitation mechanism is further examined via phase-space particle distributions and resonance conditions. To our knowledge, this work presents the first global electromagnetic gyrokinetic simulation and identification of the AITG mode in the low-$\beta$ regime with a reversed magnetic shear profile. This identification is particularly important, as previous global gyrokinetic simulations have often referred to AITG without providing clear and self-consistent physical criteria or sufficient diagnostic evidence to distinguish it from other electromagnetic instabilities, such as KBM or BAE.

This paper is organized as follows. Section \ref{sec:code} presents the gyrokinetic electromagnetic simulation model in TRIMEG-GKX, the ITG-KBM mode transformation benchmark against GKNET \cite{OBREJAN20178}, and the effect of ion-to-electron mass ratio on the instability. In section \ref{sec:instability}, the two-dimensional (2D) and radial mode structure of the AITG mode are described. The real frequency of this instability is analyzed, and differences compared to ITGs are highlighted. Section \ref{sec:characteristics} further explores key characteristics, including the dispersion relation and polarization properties, demonstrating why this instability can be identified as an AITG mode and examining its connection to experimentally observed low frequency modes LFAM. A comparative table summarizing the characteristics of different instabilities, including BAE, KBM, KIM, and the ITG mode, is also presented in this section. Section \ref{sec:resonance} presents the distributions of different instabilities in the $E-\Lambda$ phase space and their resonance characteristics, revealing the differences in the driving mechanisms of AITG and ITG modes.  Finally, Section \ref{sec:conclusions} summarizes the simulation results for AITG mode and discusses future work.  

\section{\label{sec:code}Gyrokinetic simulation model}
In this work, we apply the collisionless global electromagnetic simulation code, TRIangular MEsh-based Gyrokinetic (TRIMEG)-Generalized Kinetics eXtended (GKX), to study instabilities driven by density and temperature gradients. The TRIMEG framework has been developed for electromagnetic gyrokinetic particle simulations based on both structured (TRIMEG-GKX) \cite{lu2021development,lu2026cpc} and unstructured meshes (TRIMEG-C1) \cite{lu2019development,lu2025high}. 
To simulate the electromagnetic instabilities with gyrokinetic/drift-kinetic electrons, the pullback mixed-variable (PBMV) scheme \cite{mishchenko2014pullback} has been implemented in the TRIMEG code. 
This scheme is one of the most efficient approaches for addressing the cancellation problem that appears in the $p_{\parallel}$ formulation and has been widely used in gyrokinetic codes, such as ORB5\cite{lanti2020orb5}, EUTEPRE\cite{kleiber2024euterpe}, TRIMEG \cite{lu2023full}, GTS\cite{startsev2024verification}, and XGC\cite{hager2022electromagnetic}. In addition to the PBMV scheme, several new features have been incorporated into the TRIMEG framework. These include the strong form treatment in the particle pusher, the filter-free treatment \cite{lu2026cpc}, the piecewise field-aligned finite element method \cite{lu2025piecewise}, and the high-order C1 finite element method \cite{lu2025high}.
\subsection{Model and equations}
Using the mixed-variable method, the parallel magnetic perturbation $\delta A_{\parallel}$ can be decomposed into two parts \cite{mishchenko2014pullback}
    \begin{equation}
        \delta A_\parallel = \delta A_\parallel^S+\delta A_\parallel ^h,
    \end{equation}
where $\delta A_\parallel^S$ and $\delta A_\parallel ^h$ represent the Symplectic and Hamiltonian part, respectively. The ideal Ohm's law is used to solve the $\delta A_\parallel^S$ 
    \begin{equation}
    \label{eq:idealOhm}
        \partial_t \delta A_\parallel^S+\partial_\parallel \delta \phi=0,
    \end{equation}
the parallel derivative is defined as $\partial_\parallel=\mathbf{b}\cdot\nabla$, $\mathbf{b}=\mathbf{B}/B$, and $\mathbf{B}$ is the equilibrium magnetic field. The $\delta \phi$ denotes the scalar electrostatic potential. $\delta A_\parallel ^h$ is used to derive the shifted parallel velocity coordinate of the gyrocenter, $u_\parallel$, 
    \begin{equation}
        u_\parallel=v_\parallel+\frac{q_s}{m_s}\langle \delta A_\parallel ^h\rangle,
    \end{equation}
where $v_\parallel$ represents the parallel velocity, $q_s$ and $m_s$ denote the charge and mass of particle $s$, respectively, $s$ indicates the particle species and $\langle ...\rangle$ represents the gyro average.

The gyrocenter equations of motion are given as follows,
    \begin{eqnarray}
        \dot{\mathbf{R}}_0 &=& u_\parallel \mathbf{b_0}^* + \frac{m\mu}{qB_\parallel^*}\mathbf{b}\times\nabla B, \\
        \dot{u}_{\parallel,0}&=&-\mu \mathbf{b_0}^*\cdot\nabla B,
    \end{eqnarray}
    \begin{eqnarray}
        \delta\dot{\mathbf{R}}  &=& \frac{\mathbf{b}}{B_\parallel^*}\times\nabla \langle \delta\phi-u_\parallel\delta A_\parallel \rangle-\frac{q_s}{m_s}\langle \delta A_\parallel ^h\rangle \mathbf{b^*}, \\
        \delta\dot{u_\parallel}&=&-\frac{q_s}{m_s}(\mathbf{b^*}\cdot\nabla\langle \delta\phi-u_\parallel\delta A_\parallel^h \rangle+\partial_t\langle \delta A_\parallel^S\rangle) \nonumber\\
        &-&\frac{\mu}{B_\parallel^*}\mathbf{b}\times\nabla B\cdot\nabla\langle \delta A_\parallel^S\rangle,
    \end{eqnarray}
where $\mathbf{b_0^*}=\mathbf{b}+(m_s/q_s)u_\parallel\nabla\times\mathbf{b}/B_\parallel^*$, $B_\parallel^*=B+(m_s/q_s)u_\parallel\mathbf{b}\cdot(\nabla\times\mathbf{b})$, $\mu=v_\perp^2/2B$, and $\mathbf{b^*}=\mathbf{b_0^*}+\nabla\langle\delta A_\parallel^S\rangle\times\mathbf{b}/B_\parallel^*$. 

Using the $\delta f$ method, the distribution function can be decomposed as $f=f_0+\delta f$, where $f_0$ is the equilibrium part and $\delta f$ is the perturbed part. The dynamic equation is as follows,
\begin{equation}
    \frac{d\delta f}{dt}=-\delta\dot{\mathbf{R}} \cdot\nabla f_0-\delta\dot{u_\parallel}\frac{\partial}{\partial u_\parallel}f_0,
\end{equation}
where the equilibrium part $f_0$ is chosen as a Maxwellian distribution, and it is assumed that $d f_0/dt=0$.

The field equations are primarily solved based on the particle and current distribution on the grid. The linearized quasi-neutrality equation with the long-wavelength approximation is as follows,
    \begin{equation}
        -\nabla\cdot\left(\sum_s \frac{q_sn_{0s}}{B\omega_{cs}}\nabla_\perp\delta\phi\right)=\sum_sq_s\delta n_{s},
    \end{equation}
where the gyrocenter density $\delta n_s$ can be calculated as $\delta n_s(\mathbf{x}) = \int d^3R \, d^3v \, \delta f_{s,v}(\mathbf{R}, \mathbf{v}) \, \delta(\mathbf{R} + \boldsymbol{\rho} - \mathbf{x})$, $\mathbf{R}$ and $\mathbf{x}$ represent the  gyrocenter position vector and particle position vector, respectively, and $\boldsymbol{\rho}$ denotes the Larmor radius vector. $\omega_{cs}=q_sB/m_s$ is the cyclotron frequency of species $s$, $\delta f_{s,v}$ can be derived from $\delta f_{s,u}$ by incorporating the parallel momentum correction induced by the electromagnetic perturbation $\langle\delta A_\parallel^h\rangle$.
    \begin{equation}
        \delta f_{s,v} = \delta f_{s,u} + \frac{q_s \, \langle\delta A_\parallel^h\rangle}{m_s} \frac{\partial f_{0s}}{\partial v_\parallel}.
    \end{equation}
Since $f_0$ follows the Maxwellian distribution as stated above, $\delta f_{s,v}$ can be further expressed as
    \begin{equation}
        \delta f_{s,v} = \delta f_{s,u} - \frac{q_s v_\parallel}{T_s} \langle \delta A_\parallel^h \rangle f_{0s},
    \end{equation}
    where $T_s$ is the temperature of species $s$.
The parallel Ampère's law is given by,
    \begin{equation}
        \begin{aligned}
            &-\nabla_\perp^2 \delta A_\parallel^{h} + \sum_s \mu_0 \frac{q_s^2}{T_s}\int \mathrm{d}^6 z\, v_\parallel^2 f_{0s}\langle \delta A_\parallel^{h} \rangle\delta(\mathbf{R} + \boldsymbol{\rho} - \mathbf{x})\\
            &=\nabla_\perp^2 \delta A_\parallel^{S}+ \mu_0 \sum_s q_s\int \mathrm{d}^6 z\, v_\parallel\, \delta f_{s,u}\delta(\mathbf{R} + \boldsymbol{\rho} - \mathbf{x}),
        \end{aligned}
    \end{equation}
where $\langle\delta A_\parallel^S\rangle$ can be obtained from Eq. \eqref{eq:idealOhm}, the ideal Ohm's law.

In this work, we employ a 4th-order Runge-Kutta time integrator. At the end of each complete time step, which consists of four sub-stages, the solution is pulled back to the symplectic form by converting $(\delta f_{s,u}, u_\parallel)$ to $(\delta f_{s,v}, v_\parallel)$,
\begin{equation}
\begin{aligned}
\delta f_{\mathrm{new}}&= \delta f_{\mathrm{old}}+ \frac{q_s}{m_s}\left\langle \delta A_{\parallel,\mathrm{old}}^{h} \right\rangle\frac{\partial f_{0s}}{\partial v_{\parallel}} \\&\xrightarrow[\;f_{0s}=f_M\;]{\text{Maxwellian}}\;\delta f_{\mathrm{old}}- \frac{2 v_{\parallel}}{v_{ts}^2}\frac{q_s}{m_s}\left\langle \delta A_{\parallel,\mathrm{old}}^{h} \right\rangle f_{0s},
\end{aligned}
\end{equation}
\begin{equation}
u_{\parallel,\mathrm{new}}= u_{\parallel,\mathrm{old}}- \frac{q_s}{m_s}\left\langle \delta A_{\parallel,\mathrm{old}}^{h} \right\rangle ,
\end{equation}
\begin{equation}
\delta A_{\parallel,\mathrm{new}}^{s}= \delta A_{\parallel,\mathrm{old}}^{s}+ \delta A_{\parallel,\mathrm{old}}^{h},
\end{equation}
where $v_{ts}=\sqrt{2T_s/m_s}$ denotes the thermal velocity of species $s$.
 
\subsection{Benchmark of ITG-KBM transition}
\begin{figure}[htbp]
    \centering
        \includegraphics[width=0.48\textwidth]{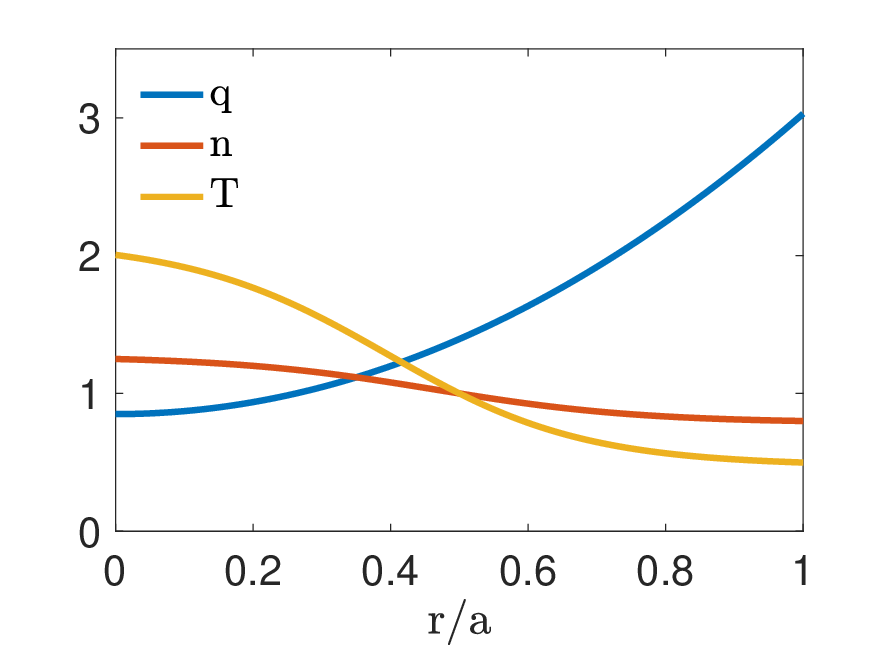}
        \caption{The CBC equilibrium profiles of safety factor $q$, density $n$ and temperature $T$.}
        \label{fig:fig1}
  \end{figure}

We perform a benchmark of the transition between ITG and KBM modes with GKNET \cite{OBREJAN20178, ZhihaoQIN2018} using the Cyclone Base Case (CBC) parameters \cite{REWOLDT2007775} as follows: the major radius $R_0=1.67 \;m$, $a/R_0=0.36$. The $q$, $n$ and $T$ profile are shown in Fig.~\ref{fig:fig1}, where 
$q=0.85+2.18(r/a)^2$, $n=\exp\{-\Delta r/L_{ns} \tanh[(r-r_0)/\Delta r]\}$, $T_s=\exp\{-\Delta r/L_{ts} \tanh[(r-r_0)/\Delta r]\}$ with $r_0=0.5a$, $\Delta r=0.3a$, $R/L_{ni,e}(r_0)=2.22$, $R/L_{ti,e}(r_0)=6.92$. $L_{n}=-(d \ln n/dr)$ and $L_{t}=-(d \ln t/dr)$ are the density and temperature gradient lengths, respectively. The density and temperature profiles are identical for electrons and thermal ions. The ion-to-electron mass ratio is $m_i/m_e=100$. The unit for the growth rate and real frequency is $C_s/R_0$, where $C_s\equiv \sqrt{T_e/m_i}$.
  
\begin{figure*}[htbp]
  \centering
    \centering
    \includegraphics[width=.4\linewidth]{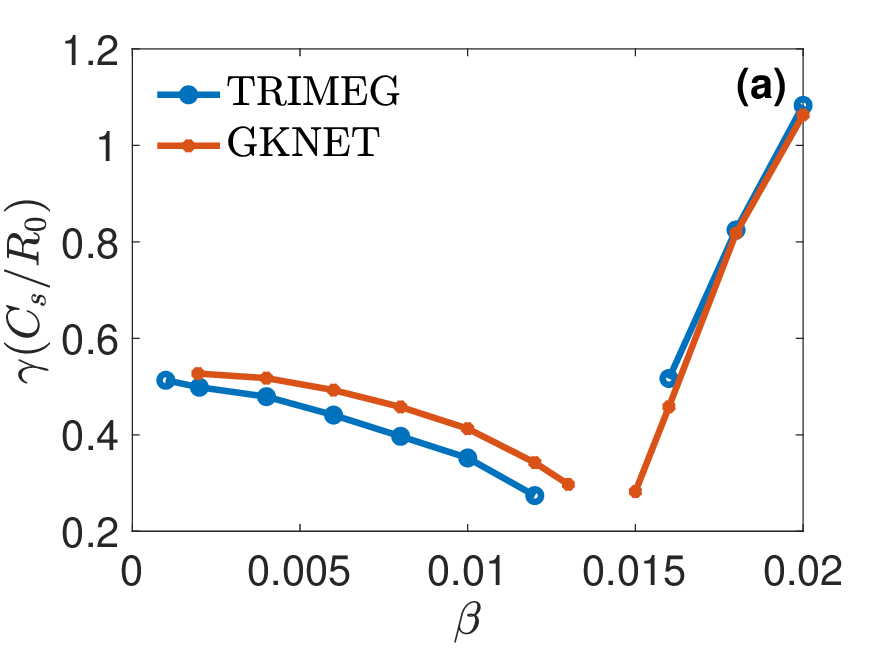}
    \centering
    \includegraphics[width=.4\linewidth]{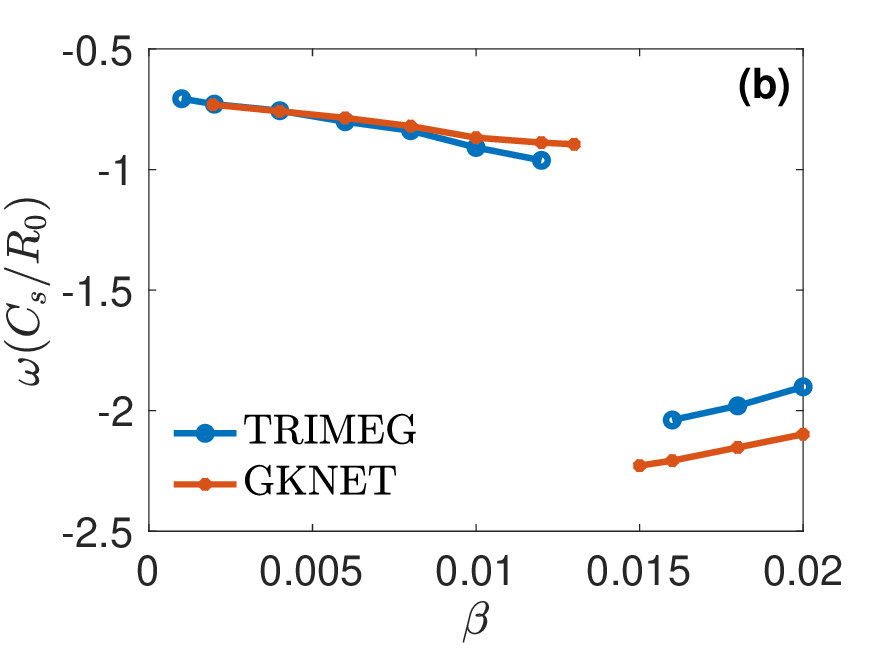}
  \caption{Growth rates $\gamma$ (panel a) and real frequencies $\omega$ (panel b) of the instabilities versus $\beta$ obtained from TRIMEG-GKX and GKNET. Red and blue lines denote the GKNET and TRIMEG-GKX results, respectively.}
  \label{fig:benchmark_beta}
\end{figure*}
Figure~\ref{fig:benchmark_beta} illustrates the transition from the ITG to the KBM branches obtained using TRIMEG-GKX and GKNET \cite{ishizawa2019global}. As shown in Fig.~\ref{fig:benchmark_beta}(a) and~\ref{fig:benchmark_beta}(b), the growth rates and real frequencies obtained from the two codes exhibit good agreement in both magnitude and trend, demonstrating the validity of the TRIMEG-GKX code. A discrepancy of approximately $\sim10\%$ is observed in the ITG mode frequency and in the KBM growth rate, while even smaller discrepancies ($<5\%$) are found in the ITG mode growth rate and the KBM frequency. These minor discrepancies are attributed to differences in the implementation of certain higher-order effects, such as gyro-averaging and the perpendicular Laplace operator.

\subsection{Effect of ion to electron mass ratio}
In gyrokinetic simulations, electron dynamics can be treated using either the hybrid kinetic–fluid model \cite{lin2001fluid} or gyro/drift-kinetic models \cite{lanti2020orb5}. In the latter case, the ion-to-electron mass ratio is typically chosen to be smaller than the realistic value, i.e., 1836, to reduce the computational cost. Previous work has shown that the use of a reduced mass ratio can introduce deviations of approximately $10\%$ in the growth rates and real frequencies for KBM\cite{ishizawa2019global} and KIM\cite{10.1063/5.0013349} compared to the realistic mass ratio. In this work, through the convergence analysis, it is found that the mass ratio not only influences the instability's frequency but also affects the radial location of the mode structure.
\begin{figure*}[htbp]
  \centering
    \includegraphics[width=.3\linewidth]{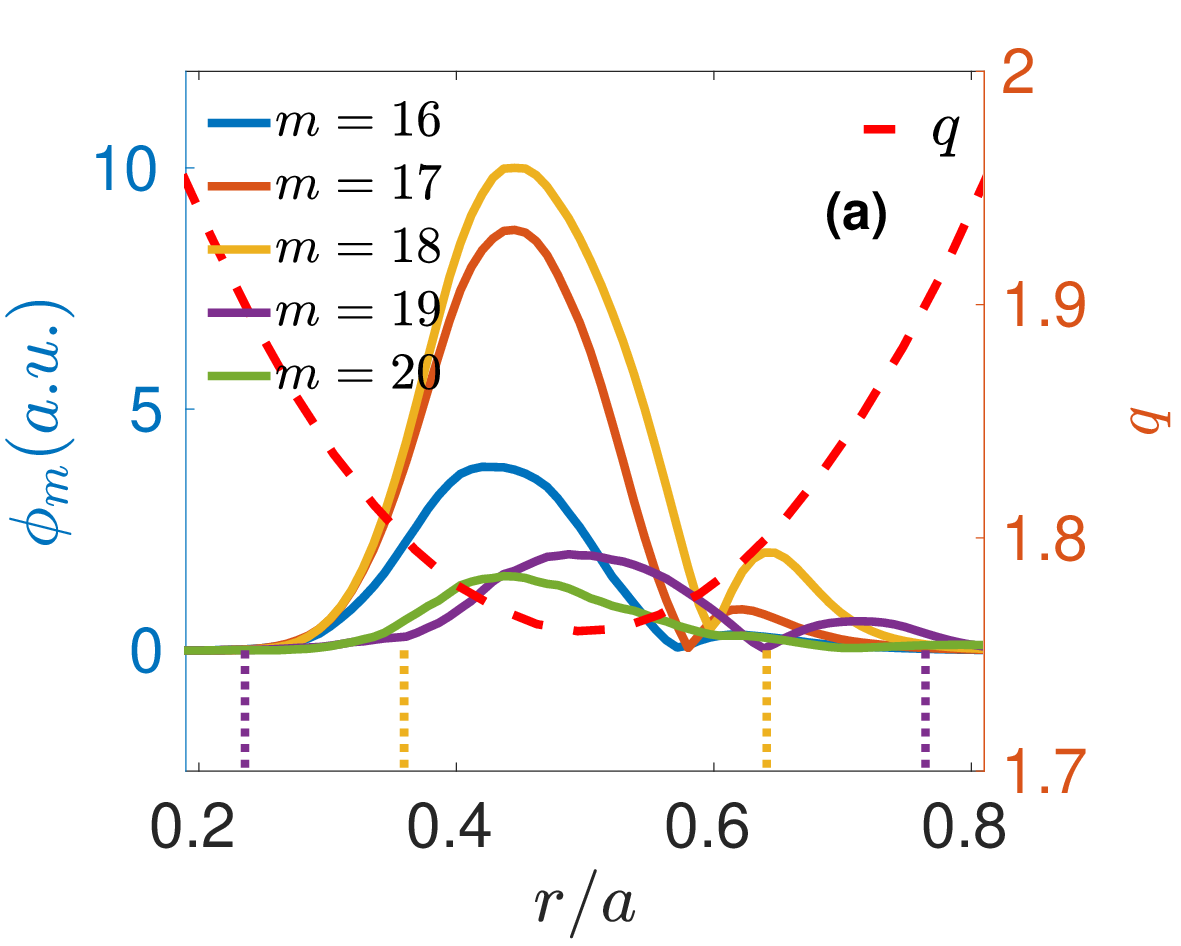}
    \includegraphics[width=.3\linewidth]{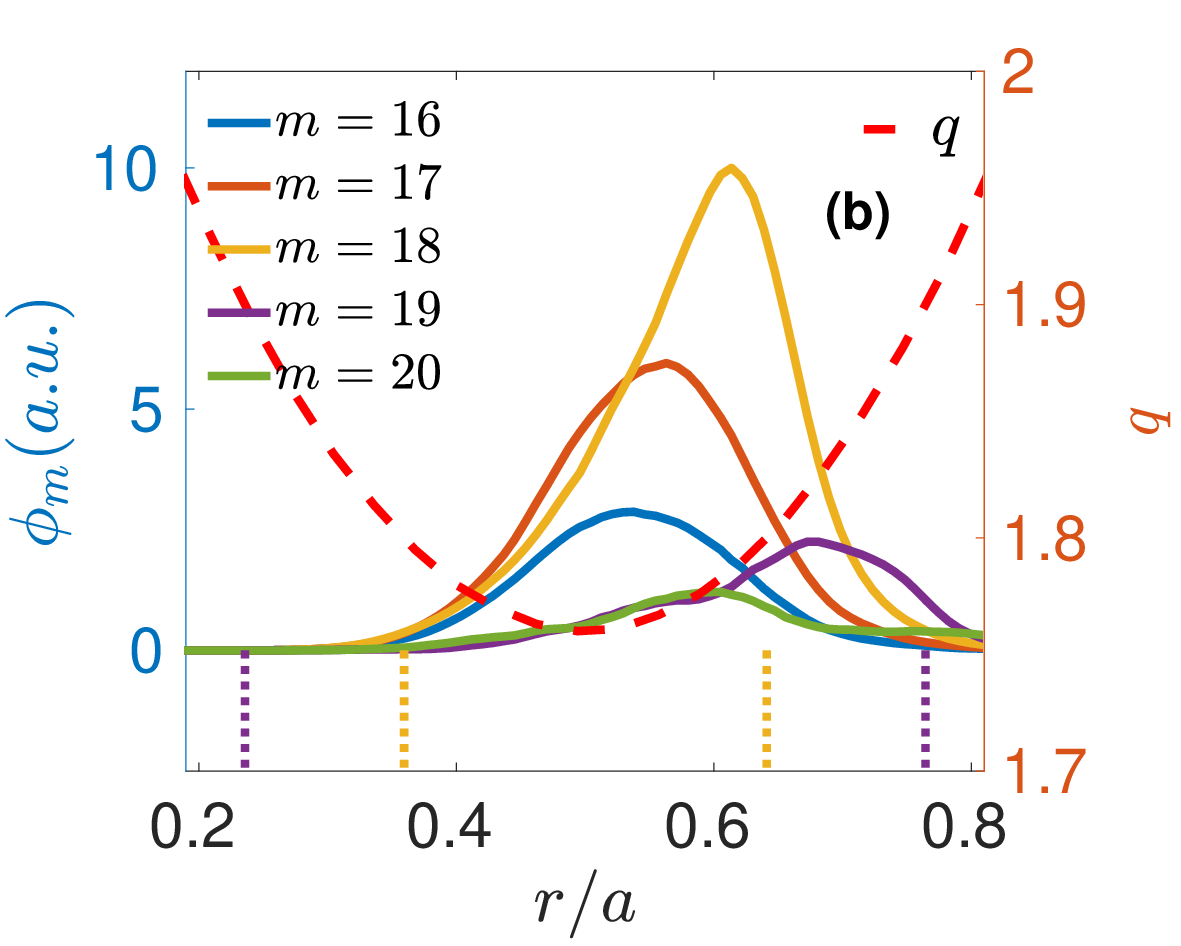}
    \includegraphics[width=.3\linewidth]{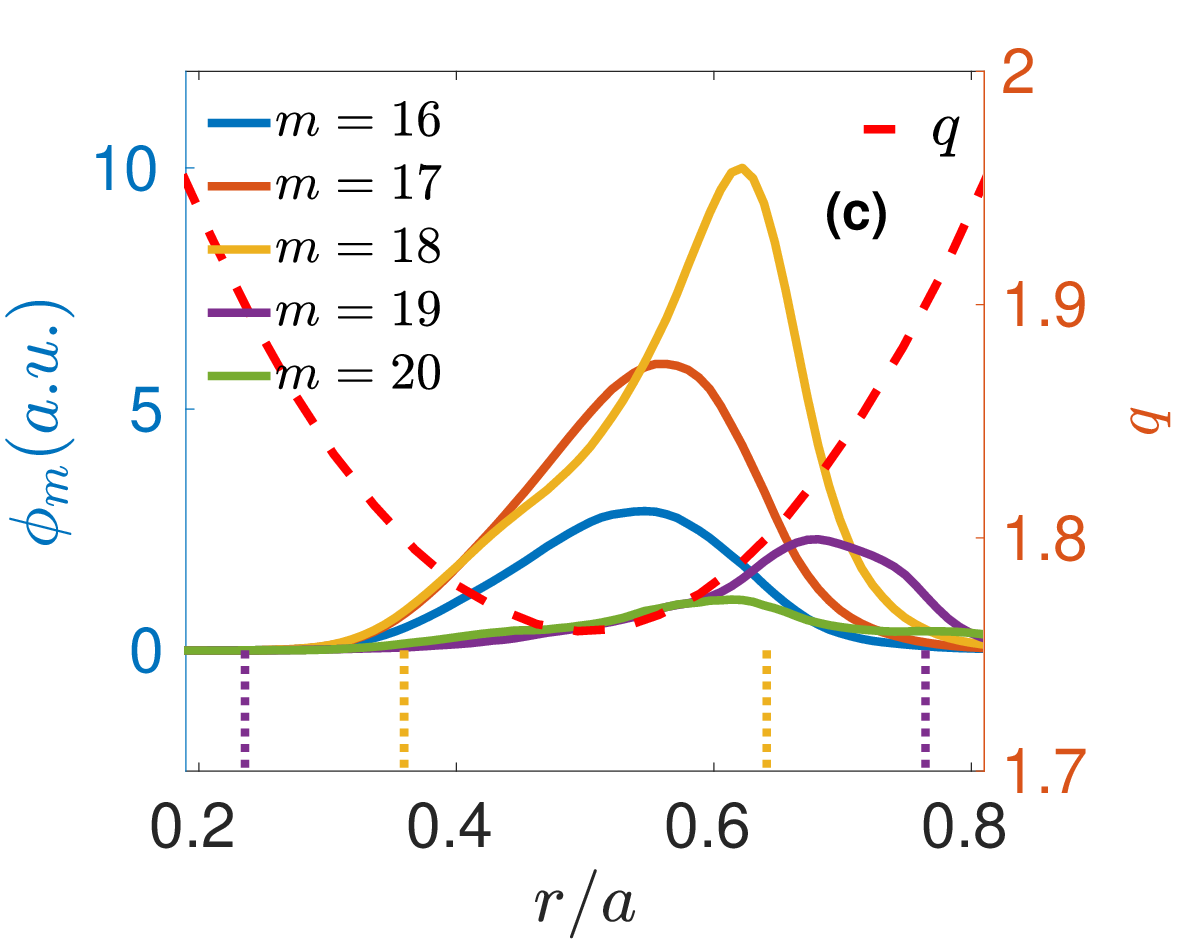}
  \caption{Radial mode structures of the instability for different ion-to-electron mass ratios with $\beta=1.8\%$. The left panel (a) corresponds to the poloidal harmonics with $m_i/m_e=100$, the middle panel (b) shows the poloidal harmonics with $m_i/m_e=400$. While the right panel (c) presents the results with $m_i/m_e=800$. }
  \label{fig:mass_ratio}
\end{figure*}

We perform simulations with reversed $q$ profiles $q(r)=1.76+2(r/a-0.5)^2$ and otherwise identical setup at a beta value of $\beta=1.8\%$ to further elucidate this effect. Figure~\ref{fig:mass_ratio} presents the radial mode structures of the KIM obtained with different mass ratios. In the figure, the yellow vertical dashed lines mark the positions of the $q = 1.8$  rational surfaces, while the purple dashed lines correspond to the $q = 1.9$ rational surfaces. For $m_i/m_e = 100$, the peak of radial mode structure shown in Fig.~\ref{fig:mass_ratio}(a) is located between the rational surface ($q=1.8$) and the zero shear position ($q_{\min}$). In contrast, for $m_i/m_e = 400$, the peak shown in Fig.~\ref{fig:mass_ratio}(b) shifts to the rational surface ($q=1.8$) closest to the zero shear position, consistent with the result obtained for $m_i/m_e = 800$ in Fig.~\ref{fig:mass_ratio}(c). Moreover, the results in Fig.~\ref{fig:mass_ratio}(b) and Fig.~\ref{fig:mass_ratio}(c) are in good agreement with those reported in Ref.~\cite{gxli_2025}. The change in mode localization likely reflects different stabilization mechanisms of the instability. Therefore, $m_i/m_e = 400$ is adopted as a compromise between computational cost and physical realism, and is used throughout the rest of the article unless otherwise stated.

\section{\label{sec:instability}TRIMEG simulation results in reversed magnetic shear}
\begin{figure*}[htbp]
  \centering
    \includegraphics[width=.4\linewidth]{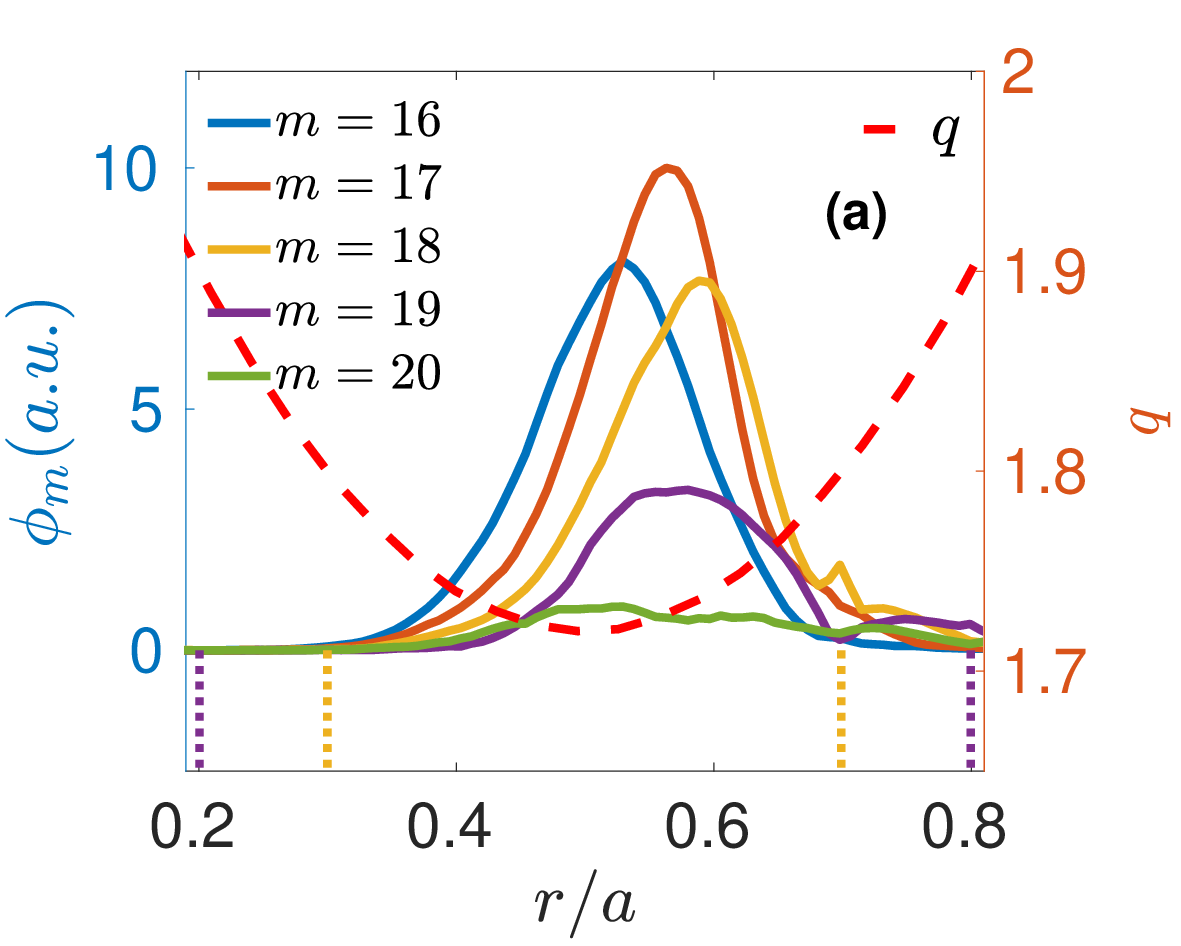}
    \includegraphics[width=.4\linewidth]{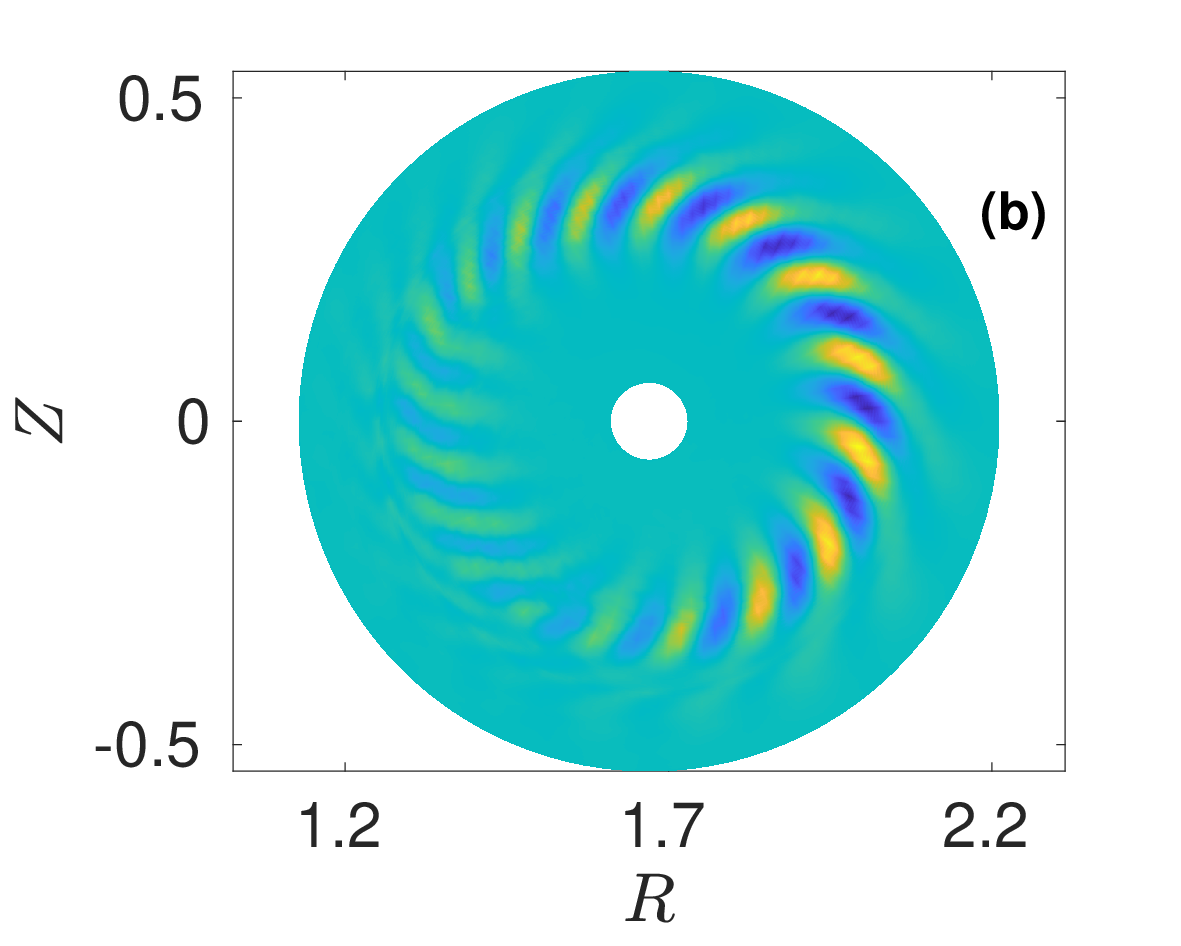}
    \includegraphics[width=.4\linewidth]{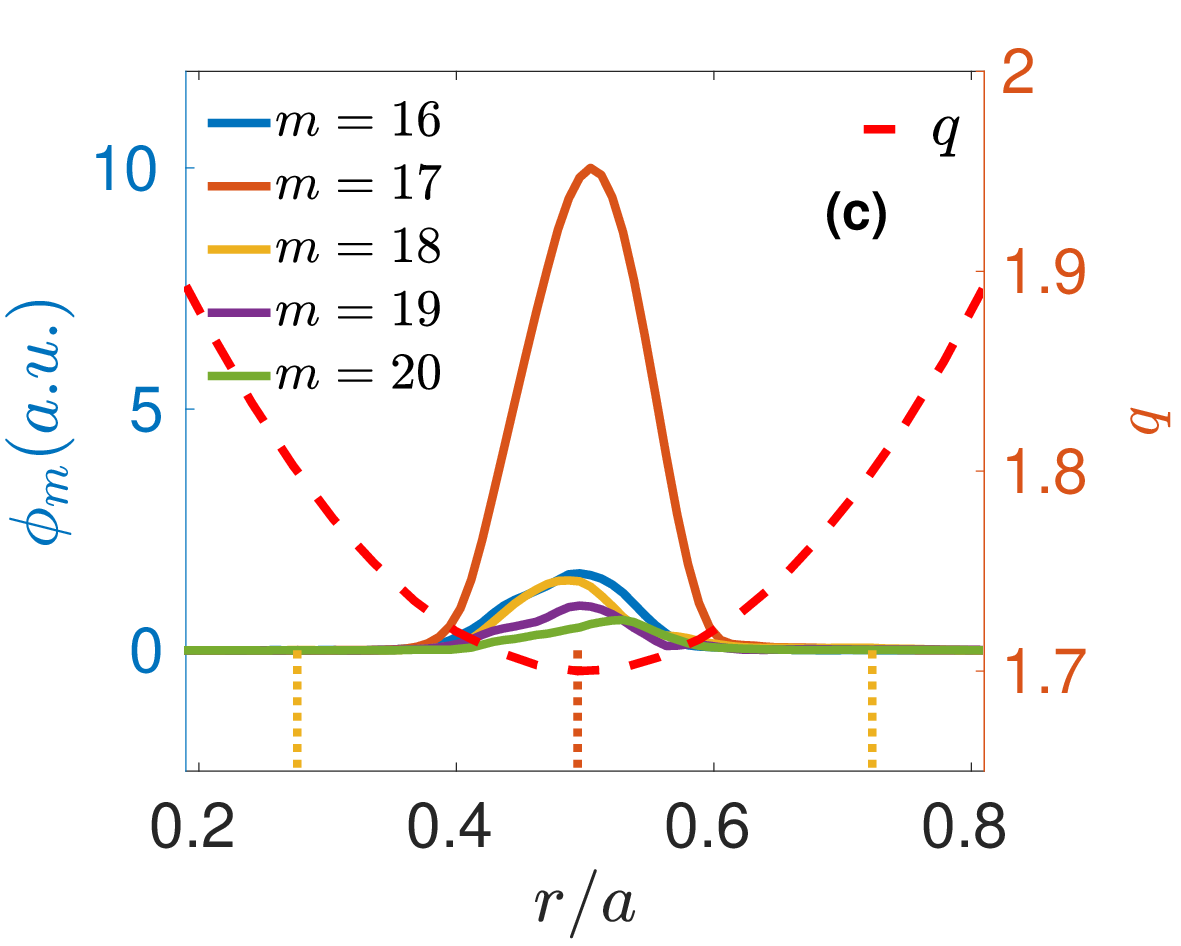}
    \includegraphics[width=.4\linewidth]{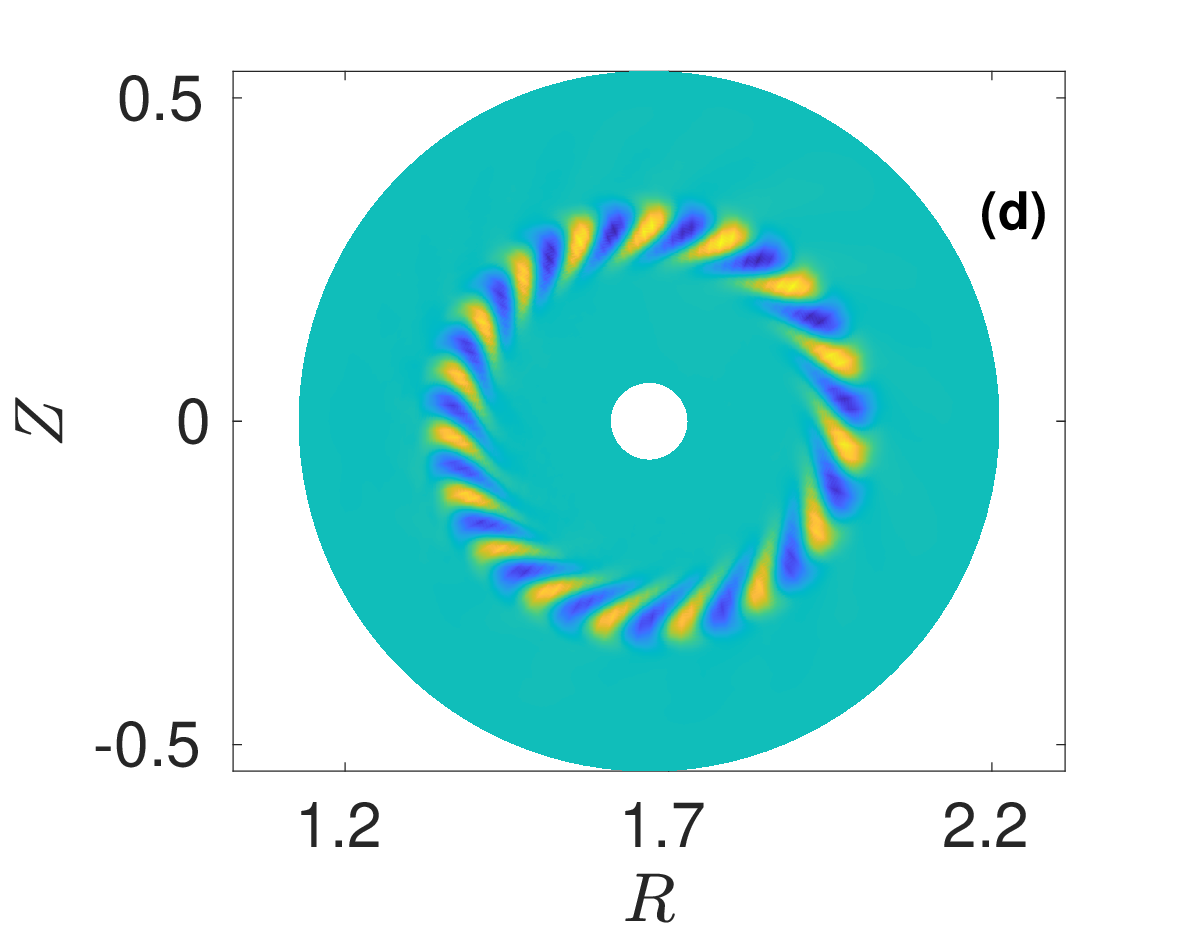}
    \includegraphics[width=.4\linewidth]{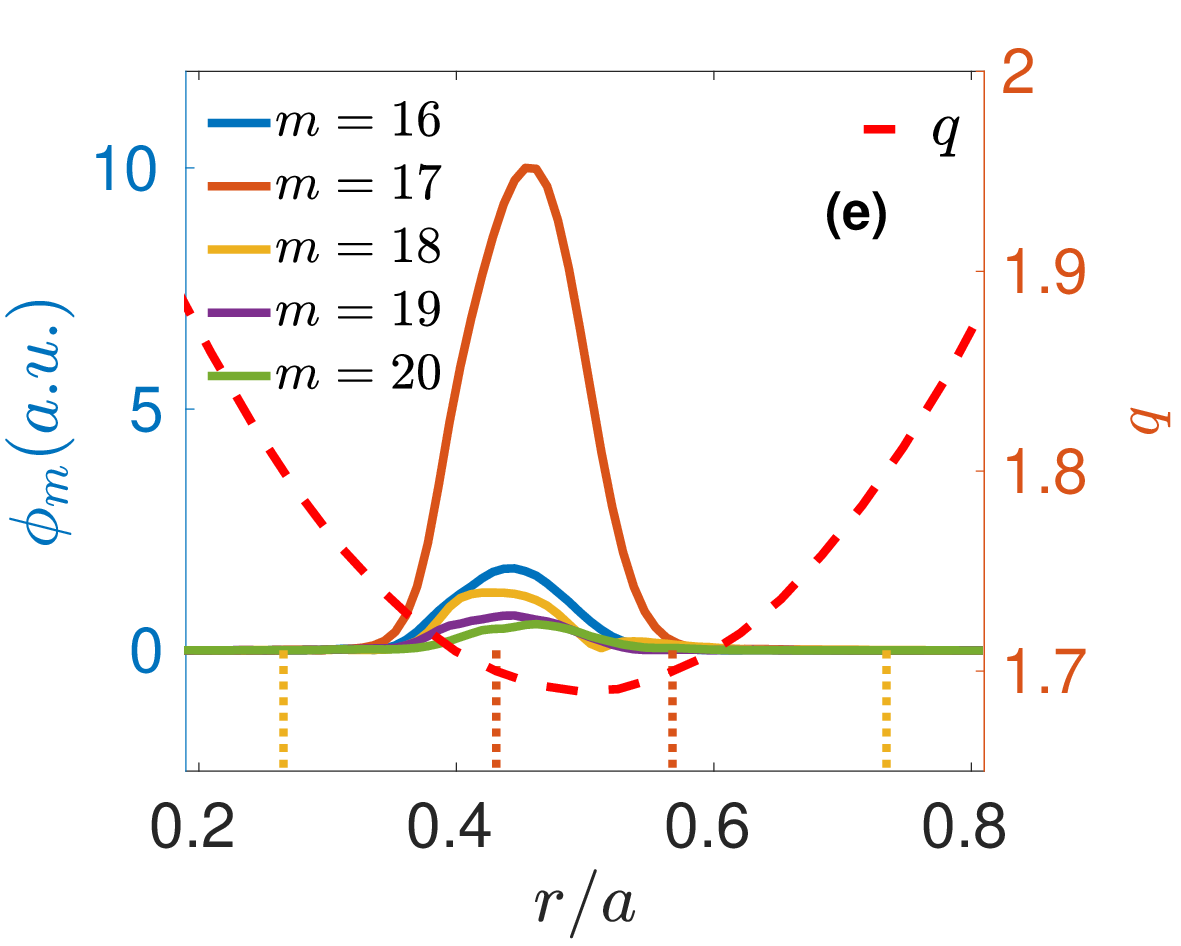}
    \includegraphics[width=.4\linewidth]{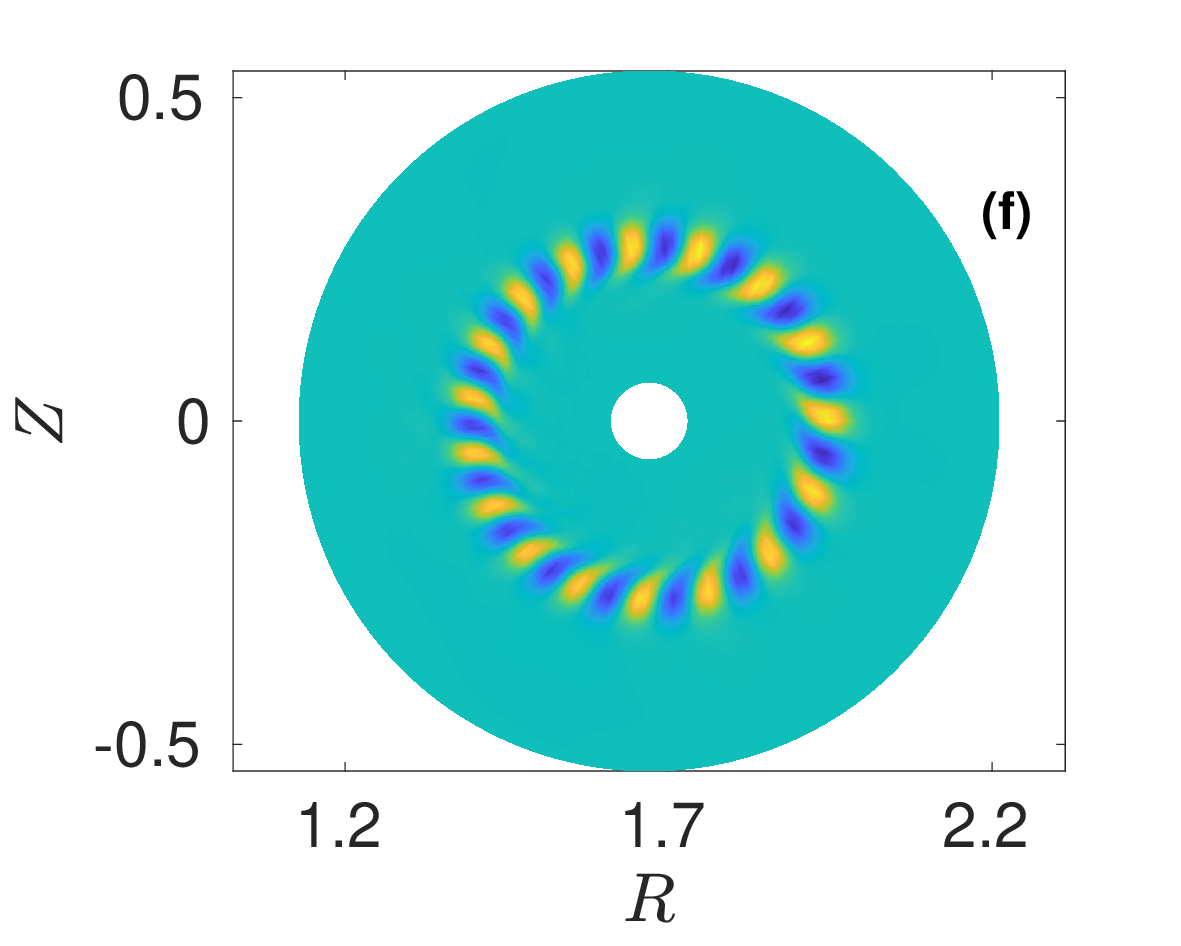}
    \includegraphics[width=.4\linewidth]{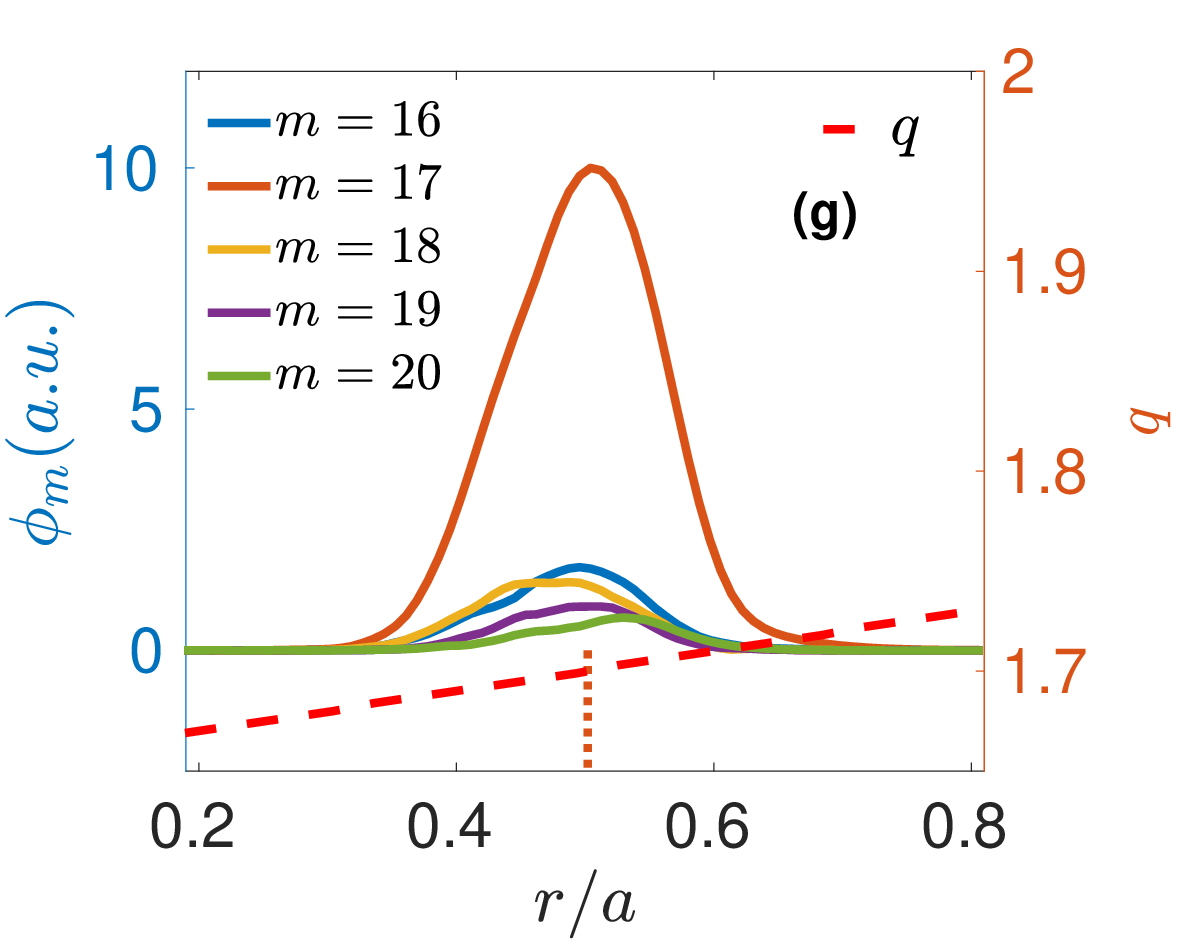}
    \includegraphics[width=.4\linewidth]{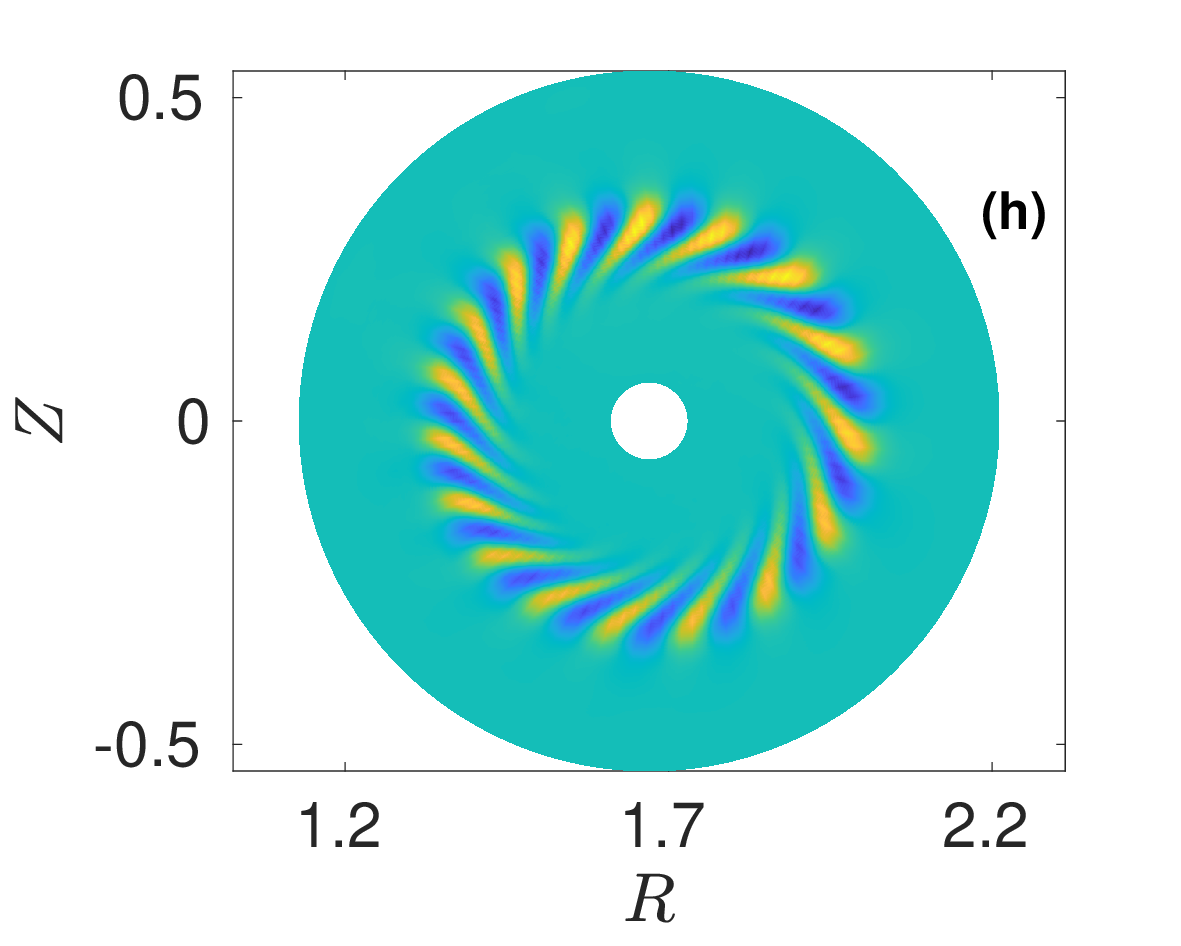}
  \caption{Poloidal harmonics (a), (c), (e), (g) and two-dimensional poloidal mode structures (b), (d), (f), (h) of the $n=10$ instabilities with different $q$ profiles. In panels (a) and (b), $q(r)=1.72+2(r/a-0.5)^2$. In panels (c) and (d), $q(r)=1.7+2(r/a-0.5)^2$. In panels (e) and (f), $q(r)=1.69+2(r/a-0.5)^2$. In panels (g) and (h), $q(r)=1.7+0.1(r/a-0.5)$.}
  \label{fig:compareAITGITG}
\end{figure*}

In this simulation, we adopt the same CBC parameters as those used in the benchmark cases except the $q$ profile. The toroidal mode number is $n=10$. The plasma $\beta$ at the diagnostic surface ($r=0.5a$) is $\beta=0.2\%$. As a typical choice, the Maxwellian distribution function is assumed for thermal ions. We first consider two different $q$ profiles, which differ only in the minimum value of the $q$ profile, $q_{\min}$. The corresponding results for mode structures are shown in Fig.~\ref{fig:compareAITGITG}(a) and~\ref{fig:compareAITGITG}(b) for $q(r)=1.72+2(r/a-0.5)^2$, Fig.~\ref{fig:compareAITGITG}(c) and~\ref{fig:compareAITGITG}(d) for $q(r)=1.7+2(r/a-0.5)^2$. 

As shown in Fig.~\ref{fig:compareAITGITG}(a), multiple poloidal harmonics of the instability are coupled with each other in the radial direction. Fig.~\ref{fig:compareAITGITG}(b) further shows that the instability is localized on the low-field side. The growth rate and real frequency of the instability are $0.1837C_s/R_0$ and $-0.688C_s/R_0$, respectively, where the negative sign corresponds to the ion diamagnetic drift direction. These characteristics collectively indicate that the observed instability is an ITG mode, which is expected under the present equilibrium conditions by previous studies\cite{REWOLDT2007775, 10.1063/1.4798392}.

In Fig.~\ref{fig:compareAITGITG}(c), this instability is dominated by a single poloidal harmonic. Fig.~\ref{fig:compareAITGITG}(d) indicates that the 2D mode structure of the instability exists over both the high and low field sides. In addition, the growth rate and real frequency of the instability are $0.3674C_s/R_0$ and $-2.829C_s/R_0$, respectively. These characteristics are markedly different from those of the ITG mode shown in Fig.~\ref{fig:compareAITGITG}(a) and~\ref{fig:compareAITGITG}(b). For convenience, the instability is referred to as the single-$m$ instability throughout the analysis.

The only difference in parameters between the cases considered in Fig.~\ref{fig:compareAITGITG}(a) and~\ref{fig:compareAITGITG}(c) is the value of $q_{\min}$, which is set to $1.7$ in Fig.~\ref{fig:compareAITGITG}(a) and $1.72$ in Fig.~\ref{fig:compareAITGITG}(c). In reversed shear configurations, such a slight variation can lead to a substantial change in the radial distribution of rational surfaces. When $q_{\min} = 1.7$, a rational surface with $m = 17$ appears at the zero shear position, i.e., $r/a=0.5$, corresponding to the toroidal mode number $n = 10$, as indicated by the vertical orange dashed line in Fig.~\ref{fig:compareAITGITG}(c). 

Figure~\ref{fig:compareAITGITG}(e) and~\ref{fig:compareAITGITG}(f) provide further evidence that the presence of a rational surface plays a crucial role in the formation of this instability for the $q$ profiles $q(r)=1.69+2(r/a-0.5)^2$. As shown in Fig.~\ref{fig:compareAITGITG}(e), when $q_{\min}=1.69$, the radial location of the instability coincides with the $n=10, m=17$ rational surface, which is slightly shifted from $r/a=0.5$. The corresponding growth rate and real frequency are $0.2842C_s/R_0$ and $-3.65C_s/R_0$, respectively. The instability exhibits a single poloidal harmonic, similar to that shown in Fig.~\ref{fig:compareAITGITG}(c) and~\ref{fig:compareAITGITG}(d). Moreover, when $q_{\min}=1.71$, no rational surface exists near this location, while for $q_{\min}=1.68$, the rational surface is located far from the zero magnetic shear position. Therefore, the instability obtained in these cases corresponds to the ITG mode. 

In contrast to the reversed $q$ profiles shown in the previous figures, the $q$ profiles $q(r)=1.7+0.1(r/a-0.5)$ in Fig.~\ref{fig:compareAITGITG}(g) and~\ref{fig:compareAITGITG}(h) are monotonically increasing with weak magnetic shear at $r=0.5a$, where a rational surface is still present. Under these conditions, the most unstable mode remains the single-$m$ instability. Based on the results shown in the above figures, the single-$m$ instability is primarily governed by weak magnetic shear and the presence of a rational surface.

The presence of such a rational surface can significantly influence wave–particle resonance effects, such as Landau damping, and is therefore likely to be one of the mechanisms responsible for the onset of this instability. To identify the instability and elucidate its underlying physical mechanisms, the characteristics are investigated in the following section. 
\section{\label{sec:characteristics}Dispersion, drive, and polarization analysis}
\begin{figure}[htbp]
    \centering
    \includegraphics[width=0.45\textwidth]{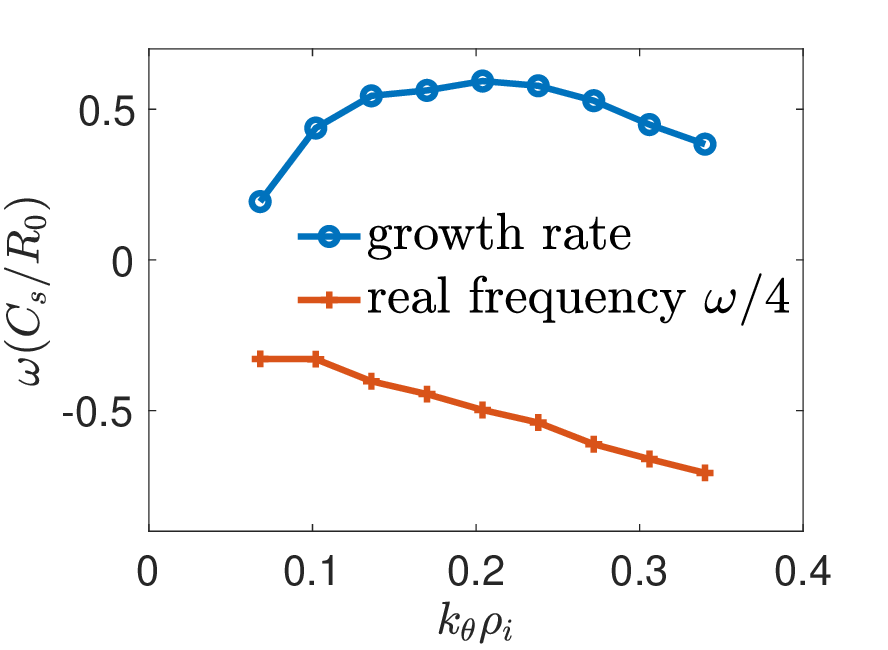}
    \caption{Real frequency (red plus symbols) and growth rate (blue circle symbols) of single-$m$ instability for varying $k_\theta \rho_i\propto n$ with $\beta=0.2\%$. The parameter $k_\theta \rho_i$ is controlled through the toroidal mode number $n$. We have chosen $q_{\min}=2$ so that $n=mq_{\min}$ is satisfied for all different values of $n$. }
    \label{fig:grfrewithn}
\end{figure}
We first investigate the dispersion relation of the instability shown in Fig.~\ref{fig:compareAITGITG}(c). The mode dispersion relation plays a crucial role in identifying the physical origin of the instability, as it reveals the dominant driving and damping mechanisms through the dependence of the frequency and growth rate on key plasma parameters. To ensure that a rational surface exists at $r/a=0.5$ for all $n$, $q$ minimal is set to $q_{\min}=2$. The corresponding results for varying $k_\theta \rho_i$ are shown in Fig.~\ref{fig:grfrewithn}, where $k_\theta \equiv m/r=nq/r$, $\rho_i=\sqrt{(2T_i/m_i)}(eB_0/m_i)$, $B_0$ is the magnetic field on axis. The growth rate represented by the blue circle line first increases and then decreases with $k_\theta\rho_i$. Additionally, the magnitude of real frequency (red plus line) increases as $k_\theta \rho_i$ increases. This dispersion relation demonstrates that the single-$m$ instability belongs to a class of kinetic instabilities related to resonances and FLR effects.
\begin{figure}[htbp]
    \centering
    \includegraphics[width=0.45\textwidth]{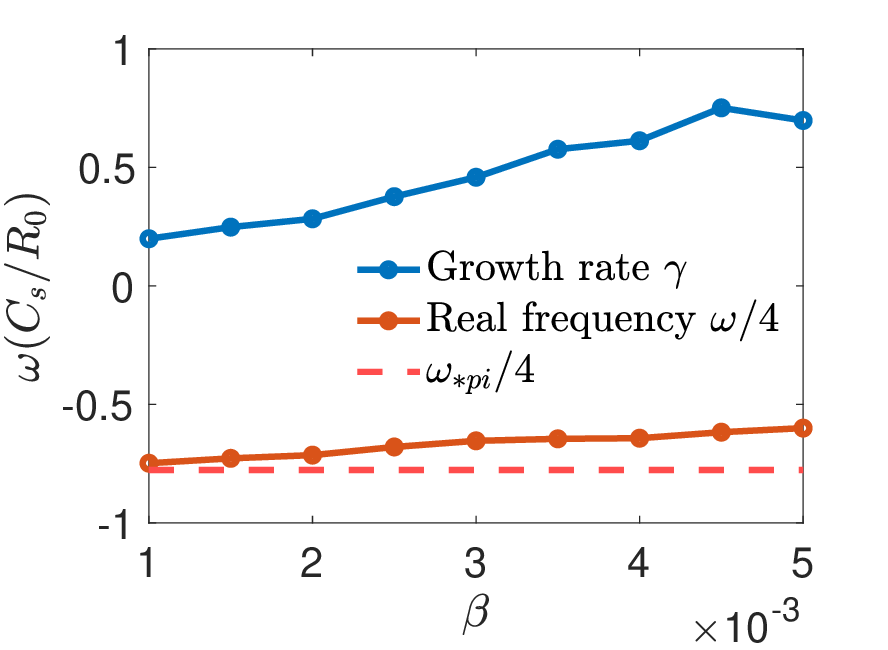}
    \caption{Real frequency (red plus symbols) and growth rate (blue circle symbols) for varying $\beta$ with $n=10$. $\beta$ is adjusted via changes in the plasma density.}
    \label{fig:grfrewithbeta}
\end{figure}

Figure~\ref{fig:grfrewithbeta} demonstrates the dependence of the growth rate and real frequency of single-$m$ instability on $\beta$ in the low-$\beta$ regime with $q_{\min}=1.7$ and $n=10$. As shown in Fig. \ref{fig:grfrewithbeta}, the growth rate indicated by the blue line increases as $\beta$ increases, indicating that the instability can be driven by pressure gradient. In contrast, for the ITG mode in the electromagnetic simulations, the growth rate decreases with increasing $\beta$ as demonstrated by the benchmark results shown in Fig. \ref{fig:benchmark_beta}, due to the coupling between the ITG mode and shear Alfv\'en waves\cite{10.1063/5.0044910}. The red curve indicates that the magnitude of the real frequency gradually approaches $\omega_{*pi}$ as $\beta$ decreases, while remaining consistently smaller than $\omega_{*pi}$ in amplitude. Here, $\omega_{*pi}=T_i/eB_0\mathbf{k} \times \mathbf{b} \cdot \nabla p_i/p_i$ denotes the ion diamagnetic frequency, where $p_i$ is the equilibrium ion pressure.

\begin{figure*}[htbp]
  \centering
    \includegraphics[width=.4\linewidth]{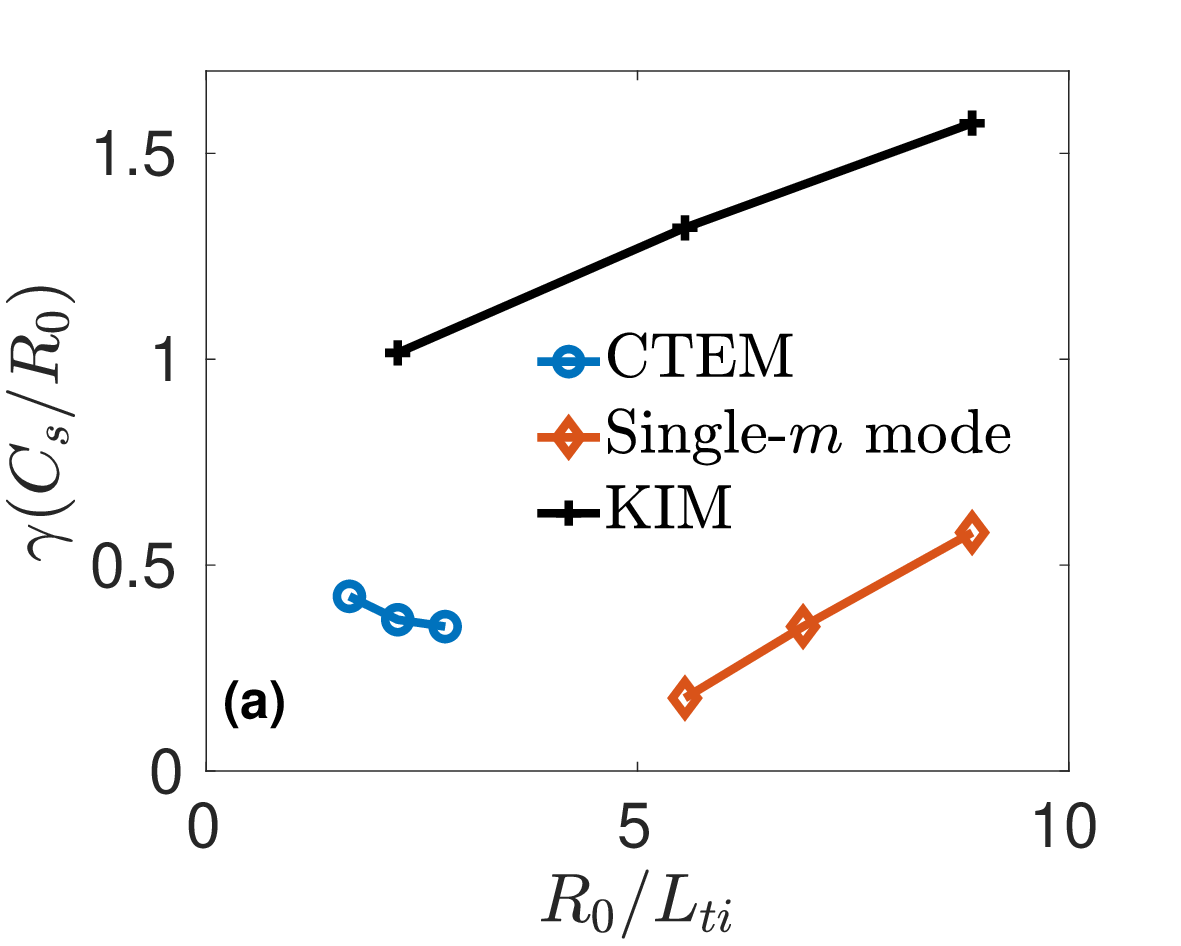}
    \includegraphics[width=.4\linewidth]{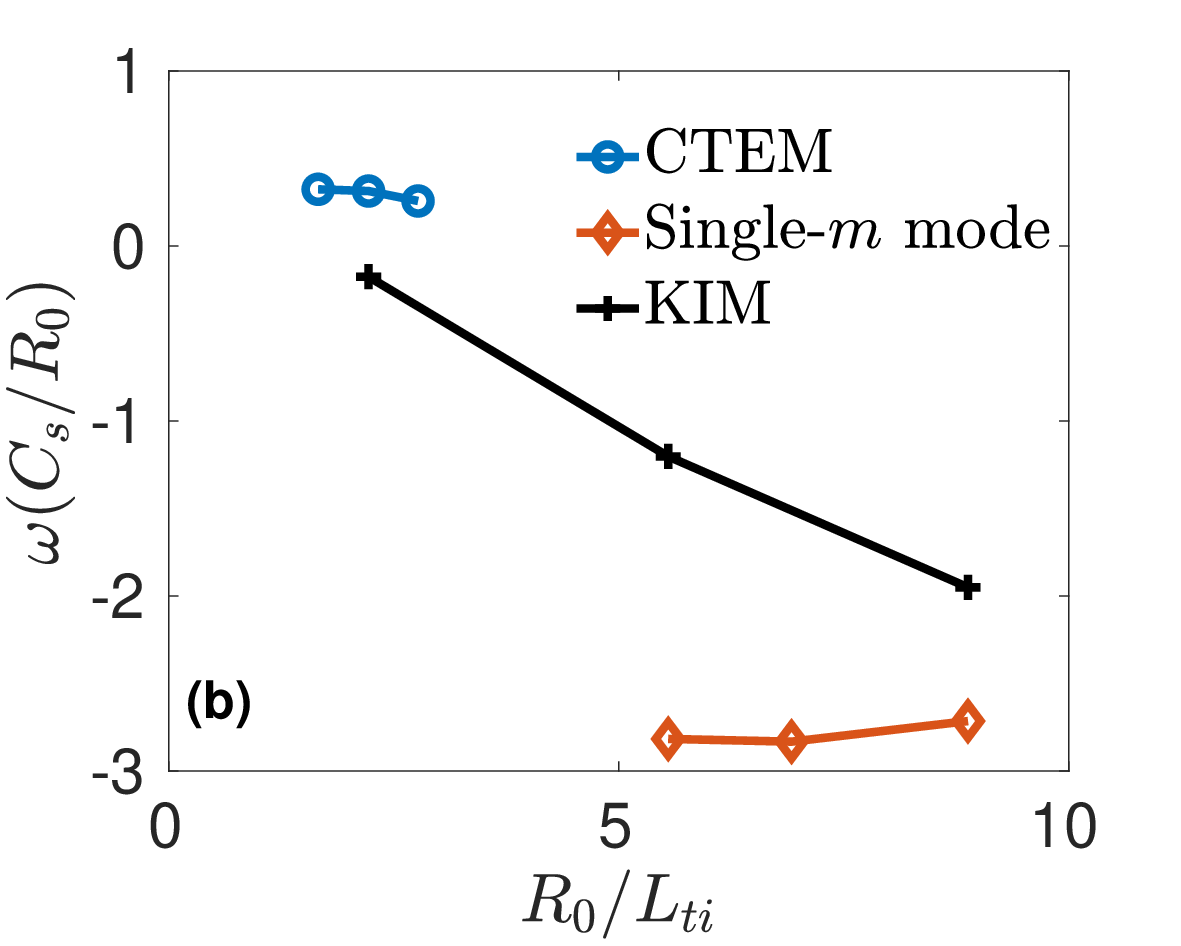}
    \includegraphics[width=.4\linewidth]{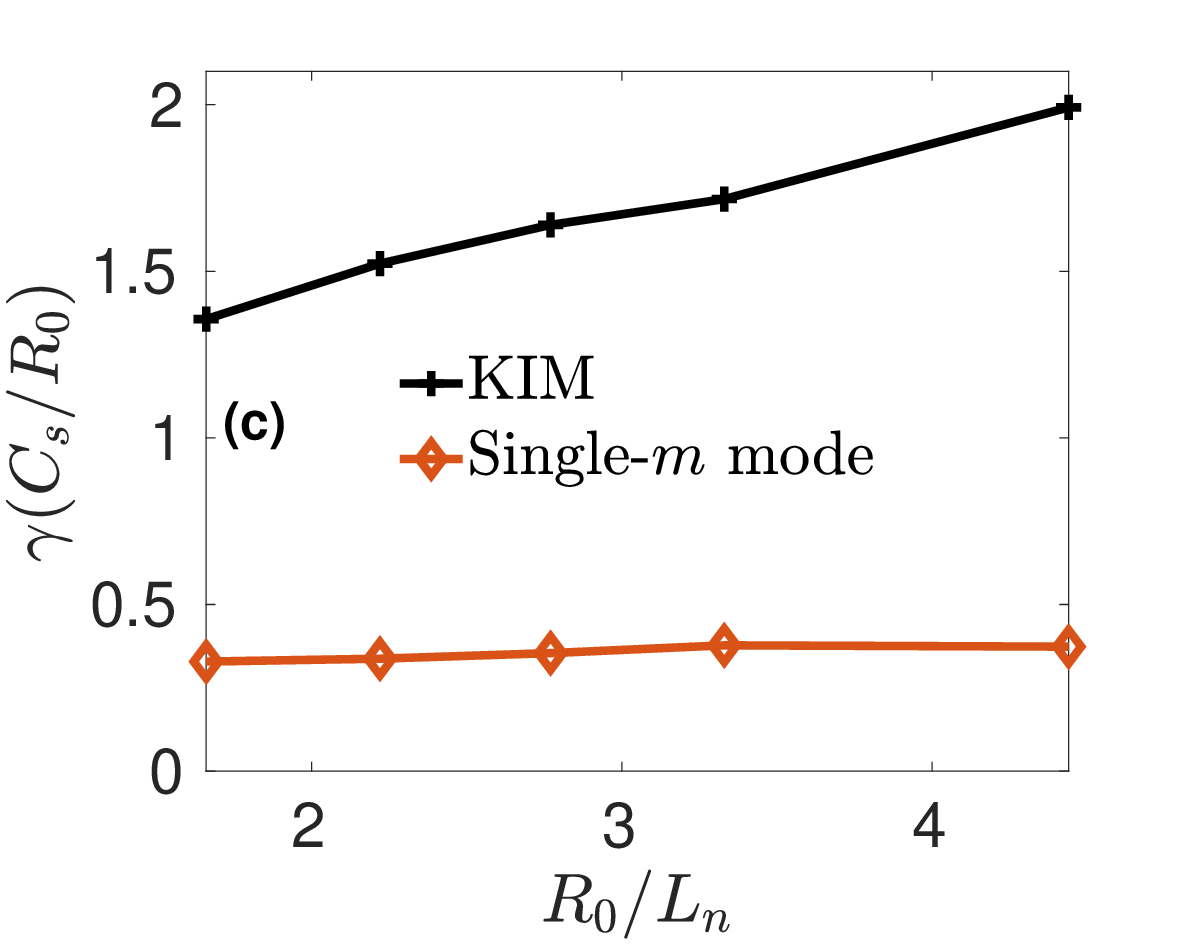}
    \includegraphics[width=.4\linewidth]{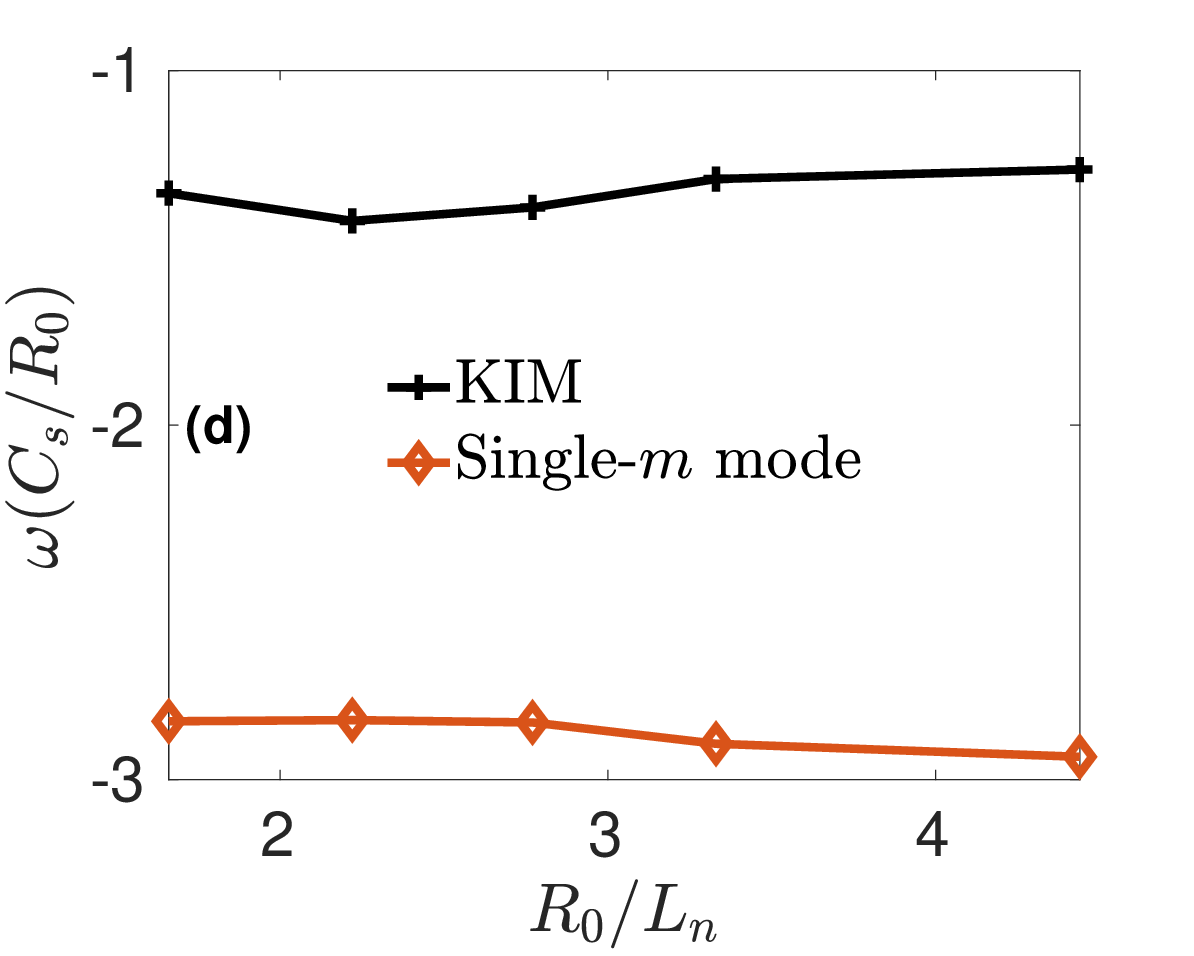}
  \caption{Real frequency (right) and growth rate (left) of  different instabilities as functions of the ion temperature (top) and density gradients (bottom). Panels (a) and (b) display the variation with $R/L_{Ti}$ at fixed $R/L_n = 2.22$, while panels (c) and (d) show the corresponding dependence on $R/L_n$ at fixed $R/L_T = 6.92$. The $q$ profile used in this figure is $q(r)=1.7+2(r/a-0.5)^2$. For single-$m$ instability and CTEM, $\beta=0.2\%$ whereas for KIM, $\beta=2\%$.}
  \label{fig:grad_scan}
\end{figure*}

In addition, the specific driving mechanism of single-$m$ instability is investigated. Its real frequency propagates in the ion diamagnetic drift direction, and no energetic particles are included in this work. Therefore, it is necessary to investigate whether this instability is driven by the ion temperature gradient or the density gradient. Figure~\ref{fig:grad_scan} shows the dependence of the real frequency and growth rate of different instabilities on the ion temperature and density gradients. As shown in Fig.~\ref{fig:grad_scan}(a), the growth rate of the single-$m$ instability (red diamonds) increases with increasing ion temperature gradient $R_0/L_{T_i}$. When the ion temperature gradient is weak, i.e., $R_0/L_{T_i}=2.22$, the real frequency in Fig.~\ref{fig:grad_scan}(b) becomes positive, indicating a transition to the collisionless trapped electron mode (CTEM), represented by the blue circles. Figure~\ref{fig:grad_scan}(c) shows the growth rates of the single-$m$ instability and KIM as functions of the ion density gradient $R_0/L_n$. The growth rate of KIM/KBM (black plus symbols) increases with increasing $R_0/L_n$, whereas the single-$m$ instability (red diamonds) exhibits only weak dependence. These results indicate that the single-$m$ instability can not be driven by the density gradient. Overall, these results demonstrate that KIM/KBM is destabilized by the pressure gradient, while the single-$m$ instability is driven exclusively by the temperature gradient.
\begin{figure*}[htbp]
    \centering
    \includegraphics[width=0.33\textwidth]{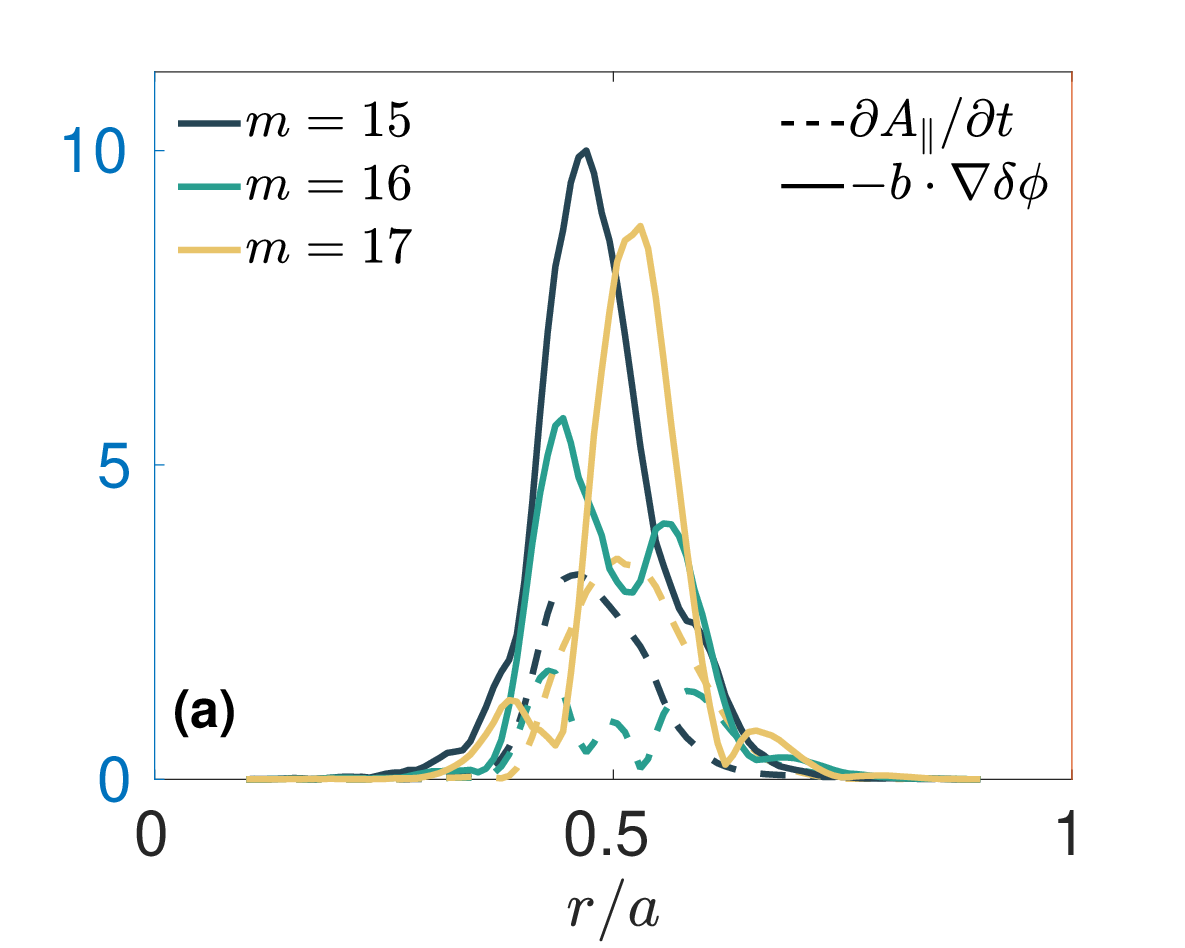}
    \includegraphics[width=0.33\textwidth]{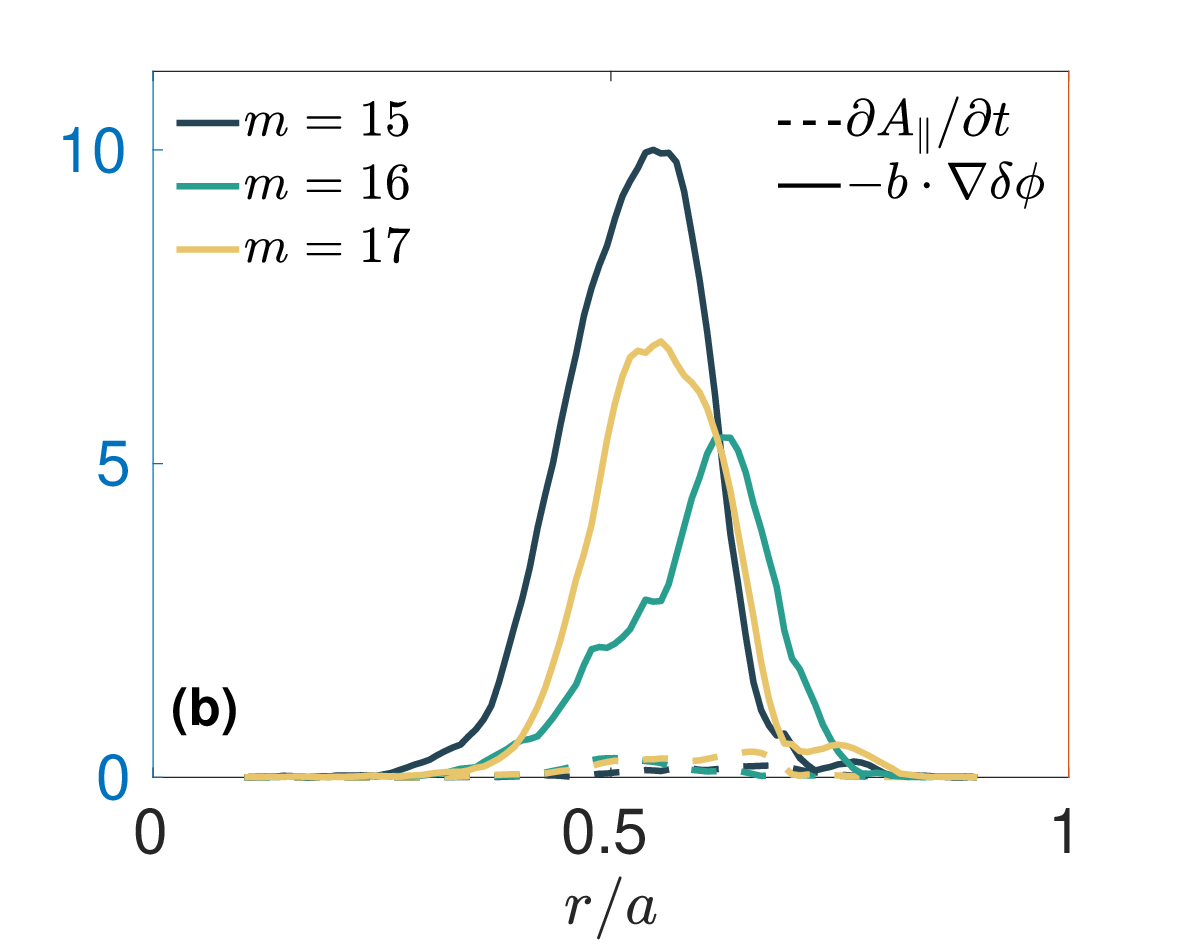}
    \includegraphics[width=0.33\textwidth]{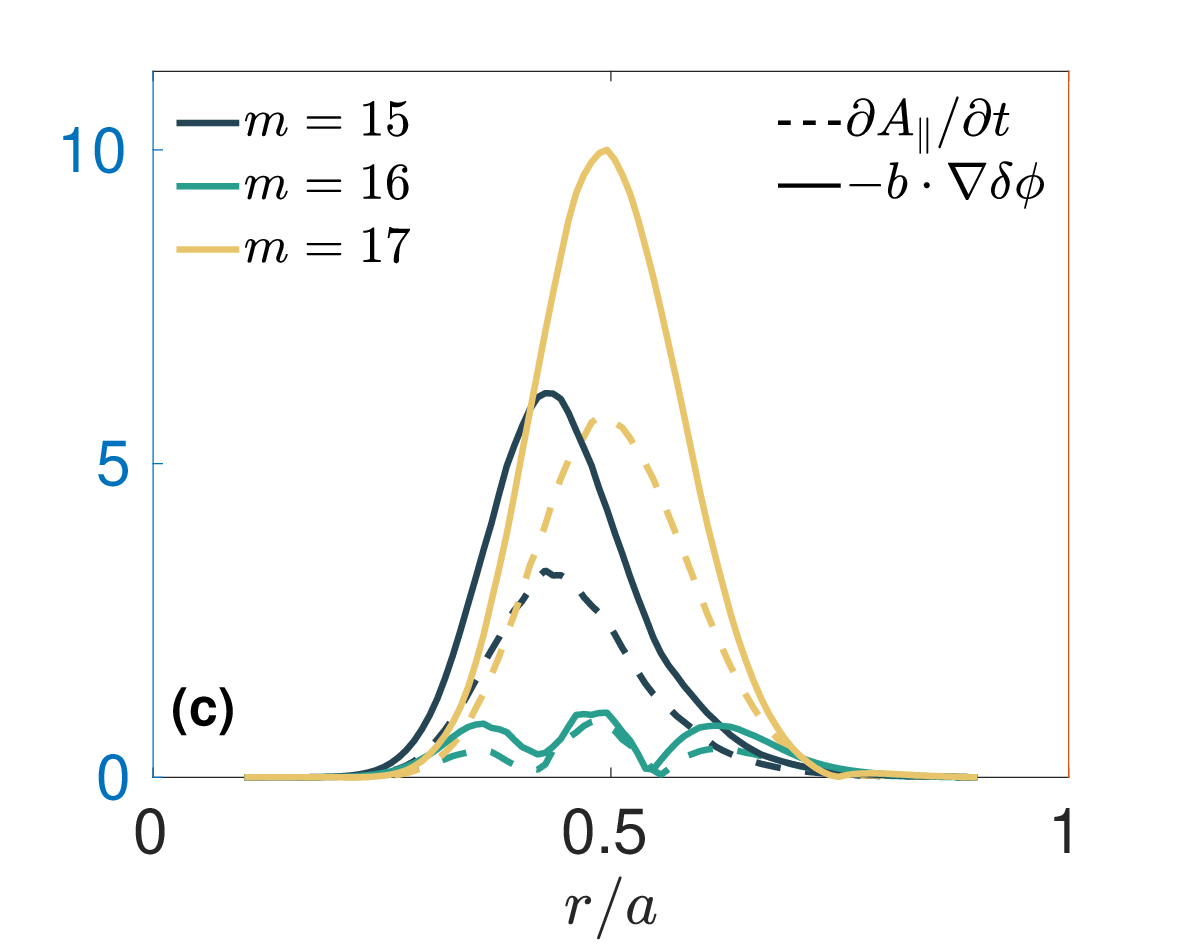}
  \caption{Polarization characteristics for different instabilities. The left panel (a) shows the polarization of the single-$m$ instability. The middle panel (b) corresponds to the ITG mode, and the right panel (c) corresponds to the KIM. The solid and dashed lines represent $-\mathbf{b}\cdot\nabla\delta\phi$ and $A_{\parallel}$, $\partial A_{\parallel}/\partial t$, respectively, while the colors denote the poloidal mode number $m$.
  }
  \label{fig:polarization}
\end{figure*}
    
Finally, the parallel electric field $\delta E_\parallel=-\mathbf{b}\cdot\nabla\delta\phi- \partial A_\parallel/\partial t$ is analyzed to examine the polarization of the instability. Comparing the two terms on the right-hand side of the equation allows us to identify whether the instability exhibits electrostatic or Alfv\'enic polarization. Figure~\ref{fig:polarization}(a), \ref{fig:polarization}(b), and \ref{fig:polarization}(c) present the polarization of the single-$m$ instability, ITG mode, and KIM, respectively. The solid line represents the electrostatic potential $-\mathbf{b}\cdot\nabla\delta\phi$, and the dashed line represents the contribution of $A_{\parallel}$, $\partial A_{\parallel}/\partial t$. As shown in Fig.~\ref{fig:polarization}(a), the polarization of this instability exhibits a finite magnetic contribution, which clearly distinguishes it from the ITG mode in Fig.~\ref{fig:polarization}(b), where the magnetic contribution is nearly zero, and shows closer similarity to that of the KBM in Fig.~\ref{fig:polarization}(c). Therefore, the instability exhibits Alfv\'enic polarization and can be classified as an electromagnetic instability.
Table~\ref{tab:instability_comparison} presents a comparative summary contrasting the instability identified in this work with other representative instabilities such as KBM, BAE, and ITG mode. The driving mechanism of single-$m$ instability is clearly different from that of the typical BAE since no energetic particles are involved. In addition, both the magnitude of the real frequency and the dependence of the growth rate on plasma $\beta$ are distinct from those of the ITG mode. Furthermore, in contrast to the KIM and KBM, the single-$m$ instability exhibits a different response to the density gradient drive $R/L_n$.
\begin{table*}[htbp]
  \centering
  \caption{Comparison of different instabilities. ``ES'' and ``EM'' denote ``electrostatic'' and ``electromagnetic'', respectively. }
  \label{tab:instability_comparison}
  \begin{tabular}{lccc}
    \toprule
    Instability & Driving mechanism & Polarization & Mode structure \\
    \midrule
    ITG
    & Ion temperature gradient & ES & Low-field side \\
    \midrule
    KIM
    & Pressure gradient & EM & Low-field side \\
    \midrule
    BAE & Energetic particles or antenna & EM & High and low-field sides \\
    \midrule
    Single-$m$ instability
    & Ion temperature gradient & Intermediate (ES--EM) & High and low-field sides \\
    \bottomrule
  \end{tabular}
\end{table*}

Based on these characteristics and previous studies, we identify the single-$m$ instability as an Alfv\'enic ion temperature gradient (AITG) mode in weak magnetic shear, namely, the weak shear Alfv\'enic-ion-temperature-gradient (\textit{WSAITG}), which is an electromagnetic instability closely related to the KBM but can exist in the low-$\beta$ regime. The real frequency of the instability is approximately 55 kHz for $n=10$. Considering the results shown in Fig.~\ref{fig:grfrewithn}, the magnitude of the WSAITG mode real frequency increases with increasing toroidal mode number $n$, in good agreement with experimental observations reported in Ref.~\cite{Chen_2016}, where frequencies in the range of 10–40~kHz were observed from the experimental measurement for toroidal mode numbers $n = 3\text{–}6$. To our knowledge, this work represents the first identification of the AITG mode with reversed magnetic shear using global gyrokinetic simulations.

WSAITG may provide a plausible explanation for the low-frequency modes observed in experiments. In DIII-D, a low frequency Alfv\'en mode (LFAM) is observed near $q_{\min}$ when the $q$ profile exhibits reversed magnetic shear and $q_{\min}$ approaches a rational value\cite{Heidbrink_2021}. Subsequent theoretical studies have investigated this LFAM in the context of both BAE and KBM, suggesting that it is not a BAE\cite{10.1063/5.0141186}. Although its real frequency is comparable to that of KBM, the nature of this instability has not been conclusively identified. While in the present work, the real frequency of WSAITG is found to be similar to that of KBM. The polarization shown in Fig.~\ref{fig:polarization}(a) and~\ref{fig:polarization}(c) indicates that WSAITG possesses an electromagnetic component, and it is weaker than that of KBM. This is consistent with experimental observations in DIII-D, where the electromagnetic signature of the low-frequency mode is weaker than that of RSAEs/BAEs\cite{Heidbrink_2021,10.1063/1.4960056}. In addition, WSAITG exhibits strong sensitivity to $q_{\min}$, in agreement with the experimental observation that the instability appears only when $q_{\min}$ is close to a rational surface. Based on these observations, we propose that the experimentally observed low-frequency mode LFAM is likely to be WSAITG.


\section{\label{sec:resonance}Resonance analyses in phase space}
The wave-particle resonance in phase space is analyzed by tracking the test particles and calculating their characteristic frequencies in terms of the transit frequency and the bounce frequency. The perturbed particle distribution, $\delta f$, can be represented in $(E,\lambda)$ space, where $E=v^2/2$ and $\lambda=\mu B_0/E$. Since $\delta f$ is a periodic function on the flux surface, integrating $\delta f$ over the full phase space yields only its zonal component, i.e., the $(m=0, n=0)$ mode. Therefore, $|\delta f|^2$ is employed to investigate the phase-space distribution of perturbed particles. In the particle statistics, the Jacobian is implicitly included in $|\delta f|^2$, where the Jacobian is $J_{E,\lambda} = 2\pi E B /(v_\| B_0)$. Consequently, the deposited marker variable is divided by the Jacobian to recover a more physically meaningful phase-space representation of the distribution function. Marker contributions are linearly deposited onto the $(E_i,\lambda_j)$ grid to obtain $\delta \tilde{f}(E_i,\lambda_j)$, with $i=1,2,\ldots,N_E$ and $j=1,2,\ldots,N_\lambda$. Here, $N_E$ and $N_\lambda$ denote the numbers of grid points in the $E$ and $\lambda$ directions, respectively.
\begin{figure*}[htbp]
  \centering
  \begin{overpic}[width=0.45\textwidth]{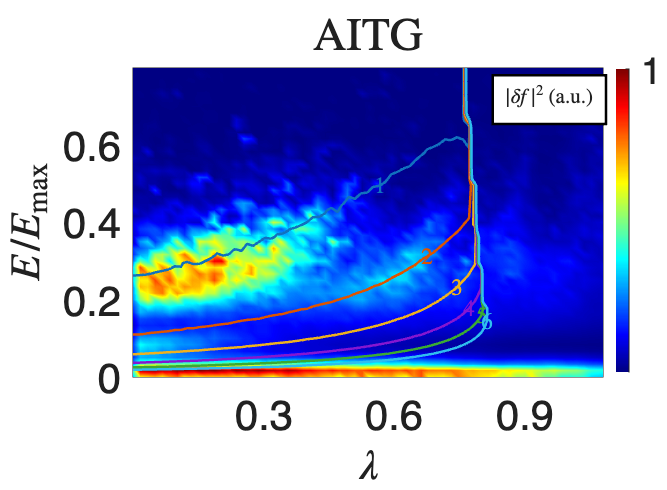}
    \put(25,67){\Large (a)}
  \end{overpic}
  \begin{overpic}[width=0.45\textwidth]{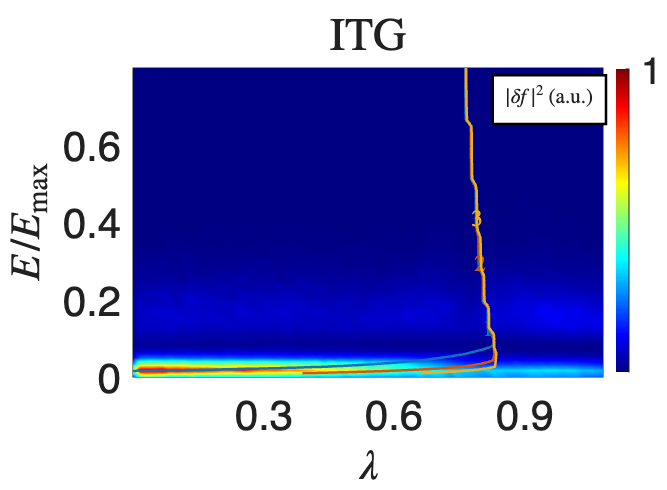}
    \put(25,67){\Large (b)}
  \end{overpic}
  \begin{overpic}[width=0.45\textwidth]{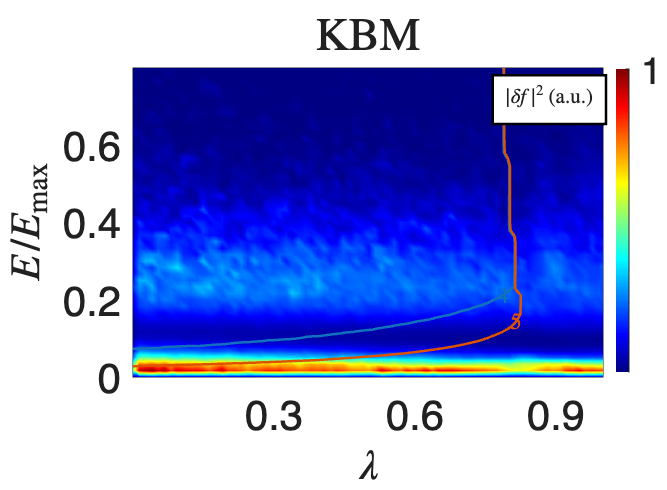}
    \put(25,67){\Large (c)}
  \end{overpic}
  \begin{overpic}[width=0.45\textwidth]{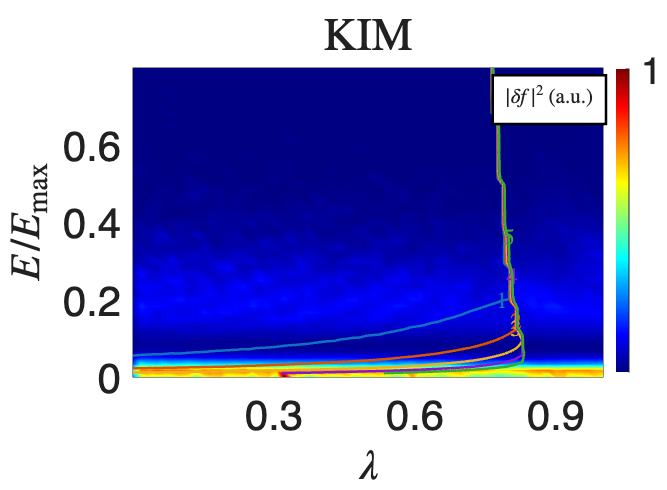}
    \put(25,67){\Large (d)}
  \end{overpic}
  \caption{Wave--particle resonance structures in phase space. Panels (a)--(d) correspond to the instability AITG, ITG, KBM, and KIM, respectively. $E_{\rm max}=(4v_{\rm th})^2/2$, where $v_{\rm  th}=\sqrt{2T/m}$. The numbers labeled on the lines correspond to the integer $p$ in Eq.~\ref{eq:omega_res}. }
  \label{fig:resonance_phase_space}
\end{figure*}

The resonance condition\cite{Wei_2025,Meng_2018} can be written as
\begin{equation}\label{eq:omega_res}
    \omega-n\omega_d+p\omega_b=0,
\end{equation}
where $n$ is the toroidal mode number, $p=m+l$, $m$ is the poloidal mode number, $l$ is an integer, 
\begin{equation}
    \omega_b=\frac{2\pi}{\oint \mathrm{d}\theta/\dot{\theta}}
\end{equation}
represents the transit/bounce frequency for circulating/trapped particles, $\omega_d$ is the toroidal transit frequency defined as
\begin{equation}
\omega_d= \frac{\omega_b}{2\pi}\oint \dot{\zeta}\frac{\mathrm{d}\theta}{\dot{\theta}},
\end{equation}
$\omega_b$ and $\omega_d$ are obtained by tracking particle orbits in the unperturbed fields. The results for WSAITG mode, ITG mode, KBM, and KIM are shown in Fig.~\ref{fig:resonance_phase_space}. In these figures, the resonance lines are calculated from Eq.~\ref{eq:omega_res} where the number labeling each line corresponds to the integer $l$ in that equation. We first compare the resonance of WSAITG and ITG modes shown in Fig.~\ref{fig:resonance_phase_space}(a) and~\ref{fig:resonance_phase_space}(b). The results indicate that, for the WSAITG instability, in addition to the particles at lower energies that resonate in a manner similar to ITG mode, there exists an additional population of circulating particles with relatively large parallel velocities participating in the resonance with $l=1$ resonance line. This population provides an extra drive for the WSAITG instability, making it the most unstable mode in this regime. These particles mainly arise because the parallel wavenumber $k_\parallel$ becomes very small near the rational surface, such that particles with sufficiently large $v_\parallel$ are required to satisfy the resonance condition. Fig.~\ref{fig:resonance_phase_space} indicates that the WSAITG and ITG modes exhibit similar wave-particle resonance in the low energy region, while the WSAITG mode receives additional energy from wave–particle resonant interactions. KBM and KIM exhibit similar wave-particle resonance patterns in phase space, which are different from those of the WSAITG mode. These features further indicate that KBM represents a different type of instability compared with WSAITG mode.

\section{\label{sec:conclusions} Conclusions and future work}
A systematic investigation of electromagnetic instabilities, i.e., ITG, KBM, and KIM, driven by temperature and density gradients has been carried out using the gyrokinetic particle-in-cell code TRIMEG-GKX. The transition from ITG to KBM has been benchmarked against GKNET based on the Cyclone Base Case, and the good agreement obtained confirms the validity and reliability of the code for electromagnetic instability studies.

In addition, an electromagnetic instability in the low-plasma-$\beta$ regime has been identified in the plasmas with reversed magnetic shear. In the low-to-moderate $\beta$ range, when the position of $q_{\min}$ coincides with a mode rational surface, a distinct instability dominated by a single poloidal harmonic develops. The key characteristics of this single-poloidal-harmonic mode have been analyzed, providing further insight into its underlying physics and parameter dependence.

The identified instability is demonstrated to be a weak shear Alfv\'enic-ion-temperature-gradient (WSAITG) mode. Its real frequency  is significantly higher than that of the ITG mode and is comparable to those of KBM, KIM, and BAE. In the low-$\beta$ regime, its growth rate increases with plasma $\beta$, in clear contrast to the behavior of the ITG mode in electromagnetic models. Furthermore, the WSAITG mode could be exclusively driven by the ion temperature gradient, unlike the KIM and KBM, which are primarily driven by the pressure gradient. The polarization characteristics of the WSAITG mode are similar to those of the KBM, indicating that it is an electromagnetic instability. A comparison between the characteristics of WSAITG and experimentally observed LFAM suggests that WSAITG may provide a plausible explanation for LFAM.

A phase-space analysis is also performed to clarify the underlying wave–particle interaction mechanism. The resonance analysis shows that, compared with the ITG mode, a substantial fraction of deeply passing particles contributes to the WSAITG resonance. This feature is consistent with the presence of rational surfaces associated with the instability. 

In future work, we will employ experimental profiles to further investigate the relationship between WSAITG and LFAM, as well as the excitation of zonal flows by WSAITG, nonlinear saturation mechanisms, associated transport levels, and its role in the formation of internal transport barriers (ITBs).



\ack{The discussions with L. Chen, F. Zonca, Q. Yu, and A. Zocco are appreciated. 
The simulations in this work were run on the TOK cluster and the MPCDF Viper/Raven supercomputers.  
This work has been carried out within the framework of the EUROfusion Consortium, funded by the European Union via the Euratom Research and Training Programme (Grant Agreement No 101052200—EUROfusion). Views and opinions expressed are however those of the author(s) only and do not necessarily reflect those of the European Union or the European Commission. Neither the European Union nor the European Commission can be held responsible for them.\\}

\bibliographystyle{iopart-num}
\bibliography{references}

\end{document}